\documentstyle[epsf]{mn}
\begin{document}

\title[Blue/UV continuum of B3-VLA radio quasars]
{The shape of the blue/UV continuum of B3-VLA radio quasars: Dependence on 
redshift, blue/UV luminosity and radio power}

\author[R. Carballo et al.]
{Ruth Carballo$^{1,2}$, J.Ignacio Gonz\'alez-Serrano,$^1$, Chris R. 
Benn$^{3}$,  
\newauthor
Sebastian F. S\'anchez$^{1,2}$ and Mario Vigotti$^{4}$\\
$^1$ Instituto de F\'\i sica de Cantabria (CSIC-Universidad de
Cantabria). Facultad de Ciencias,
39005 Santander, Spain\\
$^2$ Departamento de F\'\i sica Moderna, Universidad de
Cantabria. Facultad de Ciencias,
39005 Santander, Spain\\
$^3$ Isaac Newton Group of Telescopes, Royal Greenwich Observatory, Apdo. de 
correos 321, 38780 Santa Cruz de La Palma, Tenerife, Spain\\
$^4$ Istituto di Radioastronomia, CNR, via Gobetti 101, 40129 Bologna, 
Italy
}

\maketitle
\begin{abstract}

We present $UBVR$ photometry of a sample of 73 radio quasars, about 80
per cent complete, with redshifts 0.4--2.8.
From these data the shape of the spectral energy distribution (SED) in the
rest-frame blue/UV is analysed, using the individual sources as well
as through broad-band composite SEDs.  The SEDs of the individual
sources are generally well fitted with power-laws, with slopes
$\alpha$ ranging from 0.4 to --1.7 ($S_\nu \propto \nu^\alpha$). Two
sources with $\alpha < -1.6$ 
were excluded from the general study for having very red SEDs, 
significantly deviating with respect to the remaining sources. 
The composite SEDs cover the range $\simeq$1300--4500 \AA\ and the
only emission feature apparent from the broad-band 
spectra is the C{\sc
iv}$\lambda$1549 line, in agreement with expectations from line 
equivalent width measuremets of radio-loud quasars from the literature. 
The shape of the composites in the
log$S_\nu$--log$\nu$ plane exhibits a break at around 3000 \AA\, where
the spectrum changes from $\alpha_{\rm blue}=0.11 \pm 0.16$ at
$\lambda>3000$ \AA\ to $\alpha_{\rm UV}=-0.66
\pm 0.15$ at $\lambda<3000$ \AA. Although the broad-band 
spectral points are expected to include some masked contamination from emission 
lines/bumps, the break cannot be explained by line/bump emission, 
and most likely reflects an intrinsic trend in the continuum. 

The continuum shape is shown to depend on redshift. For the quasars
with $z<1.2$ we find $\alpha_{\rm blue}=0.21
\pm 0.16$ and $\alpha_{\rm UV}=-0.87 \pm 0.20$, i.e.  a higher
steepening. For $z>1.2$ $\alpha_{\rm UV}$ is more flat, 
$-0.48 \pm 0.12$, and there are too few spectral points
longward of 3000 \AA\ to obtain $\alpha_{\rm blue}$ and analyse the
presence of the 3000 \AA\ break. A trend similar to that between
$\alpha_{\rm UV}$ and $z$ is found between $\alpha_{\rm UV}$ and
luminosity at 2400 \AA, $L_{2400}$, with luminous quasars exhibiting a
harder spectrum.

The data show an intrinsic correlation between $L_{2400}$ and the
radio power at 408 MHz, not related to selection effects or
independent cosmic evolution.
The correlations $\alpha_{\rm UV}$--$z$, $\alpha_{\rm UV}$--$L_{2400}$ and
$L_{2400}$--$z$ appear to be consistent with accretion disc models with
approximately constant black hole mass and accretion rates decreasing
with time.  If the trends
$L_{2400}$--$z$ and $P_{408}$--$z$ are predominantly related to a selection
bias, rather than cosmic evolution, only one of the correlations
$\alpha_{\rm UV}$--$L_{2400}$ or $\alpha_{\rm UV}$--$z$ need to be
intrinsic. 

\end{abstract}

\section{Introduction}

The study of the shape of the optical/UV continuum of quasars is an
essential tool to test the different emission models invoked to
explain the observed radiation, and understand their characteristic
parameters.  Whereas the overall quasar spectrum from the infrared to
the X-ray has approximately a power-law form, $S_\nu
\propto \nu^\alpha$, with spectral index about $-1$ , in the
optical/UV region there is a bump on top of the power-law
continuum known as the big blue bump (Elvis et al. 1987 \& 
Sanders et al. 1989). The big blue bump is generally
interpreted with a two component model, consisting of the underlying
power-law and black-body-like emission from an accretion disc
(hereafter AD) around a massive black hole (Malkan 1983, Czerny \&
Elvis 1987).  The determination of the shape and strength of the ionizing UV
continumm is essential to constrain the accretion disc parameters, for 
the modelling of the broad line region
of quasars, and other aspects such as the determination of 
the $k$-corrections for quasars, which affect the derivation
of their luminosity functions.

Empirically, the shape of the optical/UV continuum of quasars is
generally parameterized by a power law, although this is a local
approximation, with the overall shape being more
complicated. Measurements of the optical/UV continuum shape at fixed
rest-frame wavelengths have been obtained by O'Brien et al. (1988),
and more recently by Natali et al. (1998), on the basis of
low-resolution spectroscopic data. O'Brien et al. selected their
quasars from availability of $IUE$ spectra and published X-ray fluxes,
and found a mean spectral index of $-0.67\pm0.05$ for the range
1900--1215 \AA. The spectral index distribution of radio-loud and
radio-quiet quasars appeared to be similar. The authors found a small
hardening with redshift, with $\alpha$ ranging from $-0.87\pm0.07$ for
$z<1.1$ to $-0.46\pm0.05$ for $z>1.1$, and a trend with luminosity at 1450
\AA, in the sense that the more luminous quasars had harder spectra.
From a joint regression analysis including spectral index, redshift
and luminosity, the authors concluded that the dominant correlation
was between $\alpha$ and $z$, and that the trend between $\alpha$ and
luminosity was due to the correlation of both variables with
redshift. Natali et al. (1998) used a complete
sample of optically selected radio quiet quasars, 
and found that the spectra in the range 5500--1200 \AA\ showed an abrupt
change around the so called ``3000 \AA\ emission feaure'' (Wills,
Netzer \& Wills 1985), with $\alpha \simeq 0.15 $ for $\lambda>3000$
\AA\ and $\alpha \simeq -0.65$ for $\lambda<3000$ \AA.

Francis et al. (1991) analysed the composite spectrum obtained from
optical spectra of around 700 quasars from the LBQS (Hewett et
al. 1991).  The individual spectra yielded slopes ranging from 1 to
--1.5 and a median value $\alpha$=--0.3 in the range 5800--1300
\AA. Francis et al. fitted to the composite spectrum a curved
underlying continuum (a cubic spline), corresponding to
$\alpha$ $\simeq$ 0.0 for $\lambda>3000$ \AA\ and $\alpha$ 
$\simeq$ --0.6 below
this limit. The authors noted that the high-luminosity quasars had 
harder spectra than the low luminosity ones.

The samples studied by Francis et al. (1991) and
Natali et al. (1998) select quasars on the basis of the presence of UV
excess (although the LBQS includes additional independent selection
criteria), and are then biased against quasars with red colours. A similar
bias is likely present in the data of O'Brien et al. (1988), who 
selected the quasars by availability of UV and X-ray measurements. 
In this paper we present a study of the
shape of the blue/UV spectrum of quasars from the B3-VLA Quasar
Sample (Vigotti et al. 1997) selected in radio, at 408 MHz. Optical selection bisases are significantly lessened in this sample, 
allowing for a thorough investigation of the shape of their blue/UV
spectrum.  Since the B3-VLA quasars have been selected at
a low frequency, where steep-spectrum, extended emission dominates, the
sample minimizes the inclusion of core-dominated quasars, for which an
additional component of relativistically beamed optical synchrotron is
likely present (Browne \& Wright 1985).

The analysis of the blue/UV continuum of the B3-VLA quasars is based
on $UBVR$ photometry of around 70 sources, with redshifts in the range
0.4--2.8. The quasar SEDs are studied both individually and from
created composite spectra, and the dependence of the blue/UV slope 
and luminosity on redshift and radio properties is analysed. 

Near infrared $K$-band imaging of 52 quasars in this work was
presented by Carballo et al. (1998). For sixteen sources the images
revealed extended emission, most likely related to starlight emission
from the host galaxy. 

\section{The sample}
 
The B3 survey (Ficarra, Grueff \& Tomasetti 1985) catalogues sources
to a radio flux-density limit $S_{\rm 408 ~MHz}=0.1$ Jy. From the B3
Vigotti et al.  (1989) selected 1050 radio sources, the B3-VLA Sample,
consisting of five complete subsamples in the flux density ranges
$0.1-0.2$ Jy, $0.2-0.4$ Jy, $0.4-0.8$ Jy, $0.8-1.6$ Jy and
$S_{408}>$1.6 Jy, 
all mapped at the VLA. Candidate
quasar identifications (objects of any colour appearing starlike to
the eye) were sought on the POSS-I red plates down to the 
plate limit, $R
\simeq 20$, yielding a sample of 172 quasar candidates.  Optical
spectroscopy was obtained for all the candidates and 125 were
confirmed as quasars, forming the B3-VLA Quasar Sample. The sample
covers the redshift range $z=0.4-2.8$, with mean redshift $z=1.16$,
and radio powers $P_{\rm 408 ~MHz}\sim$10$^{33}$-10$^{36}$ erg
s$^{-1}$ Hz$^{-1}$ (adopting $H_0$=50 km\
s$^{-1}$Mpc$^{-1}$ and $\Omega_{0}$=1).  The sample of quasar
candidates and the final B3-VLA Quasar Sample are described in Vigotti
et al. (1997).
 
The optical incompleteness of the radio quasar sample, i.e. the
fraction of quasars fainter than the optical limit of $R\simeq 20$
mag, depends on the radio flux. From $R$-band photometry of the
complete radio quasar sample presented by Willot et al. (1998), we
infer that the fraction of radio quasars with $R>20$ is 35 per cent
for the flux range $0.4<S_{408}<1.0$ Jy (mean flux density 0.66 Jy)
and 45 per cent for $0.2-0.4$ Jy (mean flux density 0.28 Jy).  The
average $B-R$ colour of these quasars is 0.6 (similar to the value we
found for the B3-VLA quasars on Section 4.2).  For the quasars with
$S_{408}>1$ Jy we can use the distribution of $B$-band magnitudes
obtained by Serjeant et al. (1998) for the Molonglo-APM Quasar Survey
(average flux density $S_{408}\sim 1.7$ Jy).  Adopting as limit
$B=20.6$, equivalent to $R=20$ for typical radio quasar colours, the
fraction of quasars with $S_{408}>1$ Jy fainter than this limit is
around 15 per cent.

The B3-VLA Quasar Sample contains 64 quasars with $S>0.8$ Jy and
average flux density of 2.25 Jy, 32 with $S=0.4-0.8$ Jy and average
flux density of 0.53 Jy and 29 quasars with $S=0.1-0.4$ Jy and average
flux density of 0.22 Jy. Adopting for the three groups optical
incompleteness of 15 per cent, 25 per cent and 45 per cent
respectively, the estimated optical completeness for
the total sample would be around 75 per cent, improving to 80 per
cent for $S>0.4$ Jy.

The present work is based on $UBVR$ photometry of a representative
group of 73 quasars from the B3-VLA Quasar Sample. The quasars were
selected to have right ascensions in the range 7$^{h}$-15$^{h}$
and comprise the 44 with
$S_{408}>0.8$ Jy, the 23 with $0.4<S_{408}<0.8$ Jy, the 4 with
$0.3<S_{408}<0.4$ and 2 out of 15 with 0.1$<S_{408}<0.3$. The sample
is equivalent to the Quasar Sample (except for the R.A. constraint)
for $S>0.4$ Jy, and includes only a few quasars with fainter radio
fluxes, although generally close to the limit. We estimate therefore the 
optical completeness of the studied sample to be about 80 per cent.

\section{Observations, data reduction and photometric calibration}
 
The optical images were obtained on 1997 February 5$-$8, using a
1024$\times$1024 TEK CCD at the Cassegrain focus of the 1.0-m JKT on
La Palma (Spain), and on March 11$-$12, using a 2048 $\times$2048 SITe
CCD at the Cassegrain focus of the 2.2-m telescope on Calar Alto
(Spain). $U, B, V$ and $R$ standard Johnson filters were used and the
pixel scale was 0.33 arcsec pixel$^{-1}$ for the JKT and 0.533
ascsec pixel$^{-1}$ for the 2.2-m telescope. The field of view was
$\sim$6'$\times$6' for all the images. A standard observing procedure
was used. Bias and sky flat fields for each night and filter were
obtained during the twilight. Faint photometric standards
(Landolt 1992 and references therein) were observed each night in
order to obtain the flux calibration.  The exposure time was different
for each quasar and filter, ranging from 60 s to 1200 s, and was set
according to the red and blue APM magnitudes from POSS-I.  The seeing
varied from $\sim 1.8$ arcsec, on February 5, 6, 7 and March 11, to
$\sim 2.2$ arcsec on February 8 and March 12.
The data were reduced using standard tasks in the {\sc iraf} software
package
\footnote{{\sc iraf} is distributed by the NOAO, which is operated by AURA,
Inc., under contract to the NSF}. 
Flat field correction was better than $0.5$ per cent. 
Large exposure time images were cleaned from cosmic-ray hits automatically.

Instrumental magnitudes for the standard stars were measured in 
circular apertures of $13$ arcsec diameter.  The flux calibration was
obtained, as a first step, assuming no colour effects in any of the
bands. For all the nights at both telescopes this assumption was
proved to be correct for filters $B, V,$ and $R$. 
At the $U$ band, for two of the nights, we took colour effects
into account, introducing the colour term $ - k''(U-B)$. 
Table 1 lists the results of the photometric calibration showing the
extinction and colour coefficients for each night/filter together with the
rms of the fits.

All nights were photometric except the second half of February 5th, during
which we had clouds. In this part of the night we only obtained data for 3
objects in two bands. The first part of the night was photometric and
the calibration data listed in Table 1 correspond to this part of the night.

\section{Optical  photometry of the quasars}

\subsection{$UBVR$ magnitudes}

Quasar magnitudes were measured on the images using the same apertures as 
for the photometric standards. In one case (B3 0724+396) a
nearby star was included within the aperture and it was subtracted by 
modelling it with a two-dimensional PSF. Measurement errors ranged from
less than 0.01 mag to 0.9 mag, and the typical values were lower than
0.15 mag (85 per cent of the data). 
Apparent $UBVR$ magnitudes, not corrected from Galactic extinction, 
and their errors are listed in Table 2, along with the observing dates and 
the colour excess $E(B-V)$ towards each object, obtained from the $N$({\rm
H}{\sc i}) maps by Burnstein \& Heiles (1982). The redshift and radio power 
of the sources is also listed in the Table.

Some of the quasars (11) were observed with the same filter twice on
the same night or on the two runs. In these cases the difference in
magnitudes never excedeed 0.2 mag, and the 
quoted magnitudes correspond to the average value.  For three quasars (B3
0836+426, B3 0859+470 and B3 0906+430) the $U$ and $V$-band data were
obtained the second part of Feb 5th, which was cloudy. We have made a
crude estimate of these magnitudes on the basis of the photometric
standards measured over the same period, and assigned them an error of
$\sim$ 0.3 mag. These objects, however, have not been used for any
further analysis.

For the 70 quasars observed under good photometric conditions
rest-frame SEDs were built, using the zero-magnitude flux densities
from Johnson (1966; $U$,$B$ and $R$ bands) and Wamsteker (1981; $V$
band). For the Galactic extinction correction Rieke
\& Lebofsky (1985) reddening law was used.  
The corrections were lower than 0.05 mag
for 80 per cent of the sources. For the remaining
sources the corrections for the $U$ band,
which is the most affected, ranged from 0.05 to 0.56 mag, with a median value
of 0.19. Figure 1 shows the SEDs of the 70 quasars, plotted in order of
increasing redshift. The SEDs will be always referred to the
log$S_\nu$--log$\nu$ plane.

Three sources in Table 2 (B3 0955+387, B3 1312+393 and B3 1444+417, see
also Fig. 1) have magnitude errors larger than 0.3 mags in several
bands. In addition, three sources have abrupt changes in their SEDs (B3
1317+380, B3 0726+431 and B3 0922+425), probably related to intrinsic
variability or not-understood errors. These six sources will not be
considered for the analysis presented in the forthcoming sections.

Histograms of the $U$,$B$,$V$ and $R$ magnitudes (corrected for
Galactic extinction) for the 64 quasars with good photometry 
are presented on Figure 2. 
The $R$-band histogram shows that most of
the quasars have $R$ magnitudes well below the POSS-I limit of $R \simeq 20$ 
used for the quasar identifications, confirming that this limit
guaranties a rather high optical completeness.

We have indicated on Fig. 1 the wavelengths of the strongest quasar
emission lines in the studied range, like H$\beta$, Mg{\sc ii}$\lambda$2798, C{\sc iii}]$\lambda$1909, C{\sc iv}$\lambda$1549 and
Ly$\alpha$. For each spectral point on the figure we have plotted the
covered FWHM, assuming the standard FWHMs (observer frame) for the
$U$,$B$,$V$ and $R$ bands from Johnson (1966).  We infer from the
figure that about 45 per cent of the spectral points do not include
within half the filter bandpass the central wavelengths of any of the
emission lines listed above. Broader emission features, such as the
Fe{\sc ii} and Fe{\sc ii}+Balmer emission bumps in the ranges 2250--2650 \AA\ and
3100--3800 \AA, could also affect the measured broad-band fluxes.

Table 3 presents average equivalent width (EW) measurements of these
emission lines for several quasar samples 

\begin{table*}
\begin{minipage}{160mm}
\caption{Summary of photometric calibration}
\begin{tabular}{@{}lccccccccc}
Date & $k(U)$ & $k''(U-B)$&rms & $k(B)$&rms & $k(V)$&rms & $k(R)$&rms \\
\hline

5 Feb& $1.55\pm0.15$&  &0.024& & & $0.54\pm0.12$&0.009& & \\
6 Feb& $0.34\pm0.07$&&0.014& & &$0.12\pm0.10$&0.027& & \\
7 Feb& $0.54\pm0.10$&&0.051&$0.24\pm0.02$&0.012&$0.21\pm0.02$&0.020&$0.15\pm0.02$&0.015\\
8 Feb& & & &$0.28\pm0.04$&0.021& & &$0.11\pm0.01$&0.008\\
11 Mar&$0.48\pm0.05$&$-0.250\pm0.023$&0.042&$0.19\pm0.03$&0.028&$0.15\pm0.01$&0.013&$0.09\pm0.01$&0.008\\
12 Mar&$0.34\pm0.05$&$-0.163\pm0.030$&0.052&$0.14\pm0.03$&0.042&$0.13\pm0.02$&0.027&$0.06\pm0.02$&0.022\\
\hline
\end{tabular}
\end{minipage}
\end{table*}

\begin{table*}
\begin{minipage}{160mm}
\caption{Photometry of the B3-VLA quasars}
\begin{tabular}{@{}lccclclclclccc}
B3 name &$z$&log $P_{408}$ & $U$ & $\sigma(U)$ & $B$ & $\sigma(B)$ & $V$ & $\sigma(V)$ & 
$R$ & $\sigma(R)$ &Date of obs.& $E(B-V)$\\
&&ergs$^{-1}$Hz$^{-1}$&& & &  & &  & & &$U$/$B$/$V$/$R$ \\
\hline
0701+392  &1.283&34.99& 18.50&0.14  &20.11&0.17  &19.69&0.12  &18.89&0.13 &6/8/6/8 &            0.118\\
0704+384  &0.579&34.72&18.82&0.12  &19.08&0.18  &18.84&0.14  &18.48&0.14 &5/7/5/7 &            0.076\\
0724+396  &2.753&35.18&19.15&0.12  &19.93&0.15  &18.78&0.04  &18.75&0.05 &12/12/12/12 &        0.062\\
0726+431  &1.072&35.00&19.52&0.29  &20.25&0.20  &18.72&0.13  & $-$ & $-$ &6/8/6/- &            0.056\\
0739+397B &1.700&35.11&17.98&0.14  &19.01&0.07  &18.87&0.14  &18.33&0.05 &6/11/6/11 &          0.050\\
0740+380C &1.063&35.66&17.59&0.17  &18.23&0.05  &17.65&0.08  &17.51&0.04 &5/7/5/7 &            0.043\\
0756+406  &2.016&35.08&19.08&0.20  &19.80&0.09  &19.24&0.06  &18.64&0.06 &12/12/12/12 &        0.048\\
0802+398  &1.800&34.96&19.26&0.12  &20.01&0.09  &19.97&0.11  &19.70&0.15 &12/12/12/12 &        0.030\\
0821+394  &1.216&35.23&17.51&0.18  &18.21&0.08  &17.39&0.06  &17.25&0.04 &5/7/5/7 &            0.034\\
0821+447  &0.893&34.99&17.42&0.16  &18.28&0.09  &17.64&0.08  &17.56&0.06 &5/7/5/7 &            0.039\\
0827+378  &0.914&35.34&17.78&0.05  &18.37&0.05  &17.94&0.06  &17.71&0.03 &5/11/5/11 &          0.028\\
0829+425  &1.056&34.60&18.69&0.20  &19.17&0.08  &18.93&0.15  &18.47&0.05 &6/11/6/11 &          0.030\\
0836+426  &0.595&34.26&20.63&0.30* &20.02&0.17  &19.70&0.30* &19.23&0.13 &5/7/5/7 &            0.033\\
0849+424  &0.978&34.84&18.59&0.20  &19.72&0.16  &18.73&0.19  &18.85&0.12 &6/8/6/8 &            0.018\\
0859+470  &1.462&35.28&17.93&0.30* &18.12&0.04  &19.49&0.30* &18.07&0.06 &5/7/5/7 &            0.009\\
0904+386  &1.730&34.86&17.73&0.11  &18.37&0.06  &18.11&0.04  &17.77&0.03 &12/12/12/12 &        0.002\\
0906+430  &0.670&35.44&19.18&0.30* &18.96&0.09  &20.00&0.30* &18.22&0.07 &5/7/5/7 &            0.009\\
0907+381  &2.16&34.84&16.41&0.07  &17.58&0.04  &17.47&0.03  &17.22&0.03 &11/11/12/11 &        0.010\\
0910+392  &0.638&33.97&18.42&0.08  &19.22&0.05  &18.81&0.06  &18.58&0.05 &12/12/12/12 &        0.002\\
0913+391  &1.250&34.94&18.70&0.19  &19.54&0.15  &18.80&0.13  &18.84&0.12 &6/7/6/7 &            0.000\\
0918+381  &1.108&35.28&20.06&0.49  &19.81&0.14  &19.23&0.19  &18.62&0.08 &6/7/6/7 &            0.002\\
0922+407  &1.876&34.75&19.52&0.14  &20.59&0.09  &20.42&0.10  &19.76&0.11 &11/11/11/11 &        0.002\\
0922+422  &1.750&35.39&18.33&0.12  &18.92&0.08  &18.36&0.06  &17.96&0.05 &7/7/7/7 &            0.006\\
0922+425  &1.879&35.54&19.94&0.29  &21.12&0.19  &21.01&0.43  &18.95&0.07 &6/7/6/7 &            0.006\\
0923+392  &0.698&34.74&15.72&0.09  &16.89&0.03  &16.62&0.04  &16.28&0.02 &6/7/6/7 &            0.005\\
0926+388  &1.630&34.98&19.12&0.18  &19.53&0.08  &19.42&0.09  &19.29&0.09 &12/12/12/12 &        0.000\\
0935+397  &2.493&34.96&21.15&0.34  &21.34&0.13  &20.89&0.14  &20.52&0.21 &12/12/12/12 &        0.005\\
0937+391  &0.618&34.56&17.60&0.16  &18.28&0.03  &18.23&0.13  &18.04&0.05 &6/8/6/8 &            0.003\\
0945+408  &1.252&35.20&17.00&0.14  &18.05&0.03  &17.70&0.08  &17.58&0.04 &6/8/6/8 &            0.000\\
0951+408  &0.783&34.51&18.76&0.11  &19.56&0.10  &18.77&0.08  &18.92&0.05 &7/11/7/11 &          0.001\\
0953+398  &1.179&34.43&18.59&0.08  &19.56&0.06  &19.14&0.06  &18.73&0.04 &12/12/12/12 &        0.000\\
0955+387  &1.405&35.29&20.04&0.33  &21.43&0.41  &21.66&0.89  &21.36&0.34 &6/8/6/8 &            0.004\\
1007+417  &0.613&34.87&15.54&0.07  &16.35&0.02  &16.05&0.03  &16.15&0.02 &6/8/6/8 &            0.000\\
1015+383  &0.380&33.65&16.99&0.06  &17.70&0.03  &17.59&0.05  &17.39&0.05 &7/8/7/8 &            0.000\\
1020+400  &1.250&35.01&16.83&0.11  &18.24&0.08  &17.83&0.09  &17.66&0.06 &6/8/6/8 &           0.000\\
1030+415  &1.120&34.70&18.54&0.11  &19.20&0.08  &18.85&0.14  &18.32&0.05 &7/11/7/11 &          0.000\\
1105+392  &0.781&34.87&19.19&0.34  &19.91&0.08  &19.17&0.20  &18.87&0.07 &6/8/6/8 &            0.000\\
1109+437  &1.680&35.99&18.15&0.14  &18.94&0.06  &18.85&0.13  &18.42&0.05 &6/11/6/11 &          0.000\\
1111+408  &0.734&35.55&16.51&0.08  &17.39&0.03  &17.26&0.06  & $-$ & $-$ &6/8/6/- &             0.020\\
1116+392  &0.733&34.01&18.01&0.12  &18.62&0.06  &18.27&0.04  &18.09&0.04 &12/12/12/12 &        0.000\\
1123+395  &1.470&34.63&17.42&0.09  &18.47&0.05  &18.33&0.04  &17.95&0.03 &12/12/12/12 &        0.000\\
1128+385  &1.735&34.44&18.78&0.23  &19.13&0.06  &18.77&0.06  &18.31&0.03 &12/12/12/12 &        0.000\\
1141+400  &0.907&34.28&20.04&0.20  &20.25&0.07  &19.71&0.08  &19.52&0.08 &12/12/12/12 &        0.000\\
1142+392  &2.276&35.33&18.32&0.15  &19.04&0.06  &18.81&0.06  &18.68&0.05 &12/12/12/12 &        0.008\\
1144+402  &1.010&34.43&17.85&0.19  &18.82&0.04  &18.24&0.08  &17.94&0.04 &6/8/6/8 &            0.000\\
1148+387  &1.303&35.27&16.06&0.10  &17.33&0.04  &17.08&0.05  &16.75&0.03 &6/8/6/8 &            0.000\\
1148+477  &0.867&35.05&16.82&0.08  &17.53&0.03  &17.13&0.03  &16.95&0.02 &7/11/7/11 &          0.004\\
1203+384  &0.838&34.42&18.38&0.10  & $-$ & $-$  &18.24&0.05  &18.14&0.04 &7/-/7/11 &           0.000\\
1204+399  &1.530&34.47&17.26&0.08  &18.23&0.05  &18.14&0.04  &17.83&0.04 &12/12/12/12 &        0.009\\
1206+439B &1.400&35.79&17.77&0.09  &18.45&0.04  &17.97&0.04  &17.34&0.02 &7/11/7/11 &          0.000\\
\hline								     
\end{tabular}							     
\end{minipage}
\end{table*}    

\begin{table*}							     
\begin{minipage}{160mm}
\contcaption{}
%\caption{Photometry of the B3-VLA quasars (contd.)}
\begin{tabular}{@{}lccclclclclcc}					     
B3 name &$z$&log $P_{408}$ & $U$ & $\sigma(U)$ & $B$ & $\sigma(B)$ & $V$ & $\sigma(V)$ & 
$R$ & $\sigma(R)$ & Obs. date & $E(B-V)$ \\				
\hline								     
1228+397  &2.217&35.26&17.88&0.08  &18.54&0.06  &18.42&0.06  &18.18&0.07 &7/7/7/7 &            0.000\\
1229+405  &0.649&34.31&19.04&0.12  &19.39&0.10  &19.33&0.14  &19.04&0.07 &7/11/7/11 &          0.000\\
1239+442B &0.610&34.37&17.94&0.07  &18.63&0.03  &18.39&0.04  &18.14&0.03 &7,11/11/7,11/11 &    0.000\\
1240+381  &1.316&34.29&17.86&0.06  &19.28&0.05  &18.87&0.06  &18.43&0.05 &12/12/12/12 &        0.000\\
1242+410  &0.811&34.68&19.71&0.20  & $-$ & $-$  &20.02&0.15  &19.65&0.13 &6/-/6/8 &             0.000\\
1247+450A &0.799&34.71&17.05&0.07  &17.93&0.03  &17.62&0.05  &17.47&0.02 &7/11/7/11 &          0.004\\
1256+392  &0.978&34.66&19.22&0.05  &19.62&0.10  &19.12&0.13  &18.85&0.05 &7/11/7/11 &          0.000\\
1258+404  &1.656&35.93&18.32&0.09  &18.63&0.05  &18.45&0.06  &18.18&0.04 &7/11/7/11 &          0.000\\
1312+393  &1.570&35.04&21.41&0.75  &22.22&0.30  &21.97&0.59  &21.76&0.49 &11/11/6/11 &      0.000\\
1315+396  &1.560&35.06&17.75&0.15  &18.69&0.04  &18.61&0.05  &18.20&0.04 &11/11/11/11 &     0.000\\
1317+380  &0.835&34.44&19.65&0.22  &20.42&0.17  &18.98&0.10  &19.64&0.09 &7/11/7/11 &       0.000\\
1339+472  &0.502&34.47&20.13&0.49  &20.26&0.11  &19.23&0.06  &18.74&0.05 &6/11/6,12/11 &    0.006\\
1341+392  &0.768&34.61&20.68&0.29  &21.18&0.09  &20.80&0.14  &20.23&0.14 &11/11/7,11,12/11 &0.000\\
1342+389A &1.533&35.15&17.84&0.11  &18.41&0.04  &18.32&0.04  &17.97&0.03 &11/11/11/11 &     0.000\\
1343+386  &1.844&35.31&17.55&0.10  &18.10&0.04  &17.87&0.03  &17.50&0.03 &11/11/11/11 &     0.000\\
1348+392  &1.580&34.88&19.19&0.14  &19.99&0.09  &19.79&0.08  &19.61&0.09 &12/12/12/12 &     0.000\\
1355+380  &1.561&34.93&18.65&0.07  &19.48&0.04  &19.41&0.07  &18.78&0.05 &11/11/11/11 &     0.000\\
1357+394B &0.804&34.20&18.08&0.06  &18.85&0.05  &18.53&0.05  &18.31&0.04 &12/12/12/12 &     0.000\\
1416+400  &0.473&33.88&19.89&0.15  &20.37&0.05  &19.94&0.11  &19.40&0.09 &11/11/11/11 &     0.000\\
1417+385  &1.832&34.41&18.54&0.07  &19.19&0.06  &19.03&0.05  &18.65&0.06 &12/12/12/12 &     0.000\\
1419+399  &0.622&33.35&18.67&0.12  &19.00&0.06  &19.01&0.07  &18.75&0.03 &12/12/12/12 &     0.000\\
1435+383  &1.600&34.89&17.47&0.09  &18.20&0.05  &18.04&0.04  &17.78&0.04 &12/12/12/12 &     0.000\\
1444+417A &0.675&34.60&18.73&0.39  &19.42&0.21  &19.39&0.34  &19.08&0.12 &6/11/6/11 &       0.002\\
\hline
\end{tabular}								
* indicates unreliable photometry (second part of the night of Feb 5th)
\end{minipage}
\end{table*}

\vskip 2truecm

\begin{figure*}
\vskip -3truecm
\epsfxsize=16cm
\epsffile{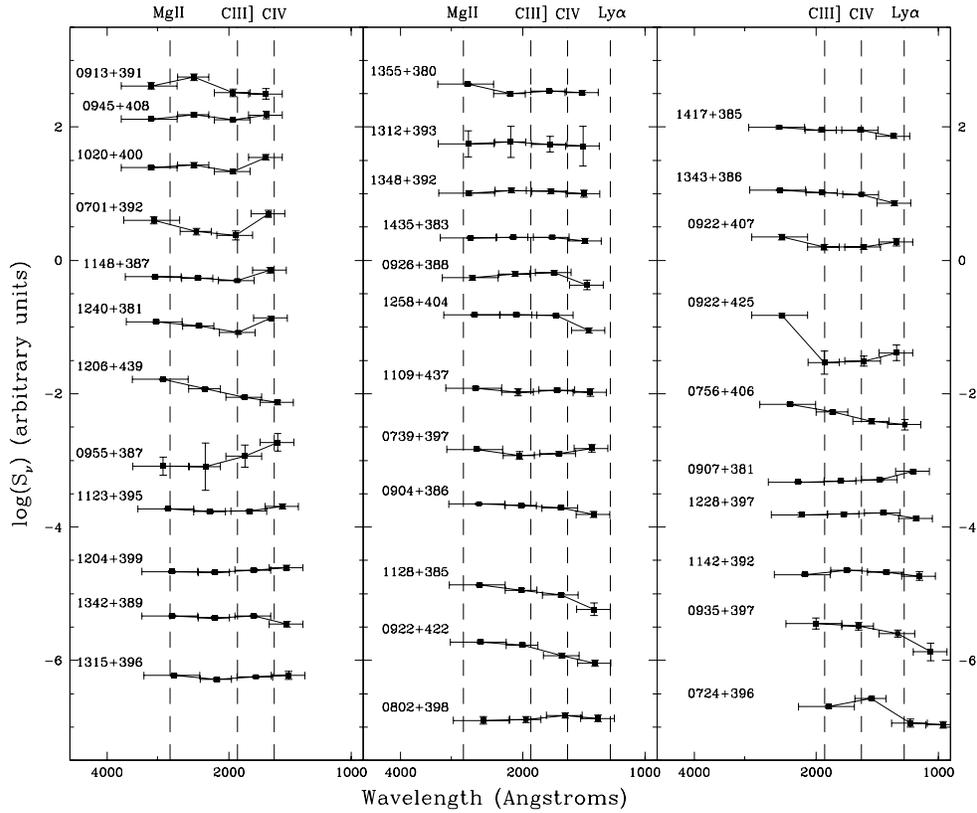}
\vskip -1.6truecm
\caption{Individual rest-frame SEDs of the 
B3-VLA quasars plotted in
order of increasing redshift. Vertical
dashed lines indicate typical broad emission lines observed in quasars:
H$\beta$, Mg{\sc ii}$\lambda$2798, C{\sc iii}]$\lambda$1909, C{\sc
iv}$\lambda$1549 and Ly$\alpha$.}
\label{}
\end{figure*}

\begin{figure*}
\epsfxsize=16cm
\vskip -3.2truecm
\epsffile{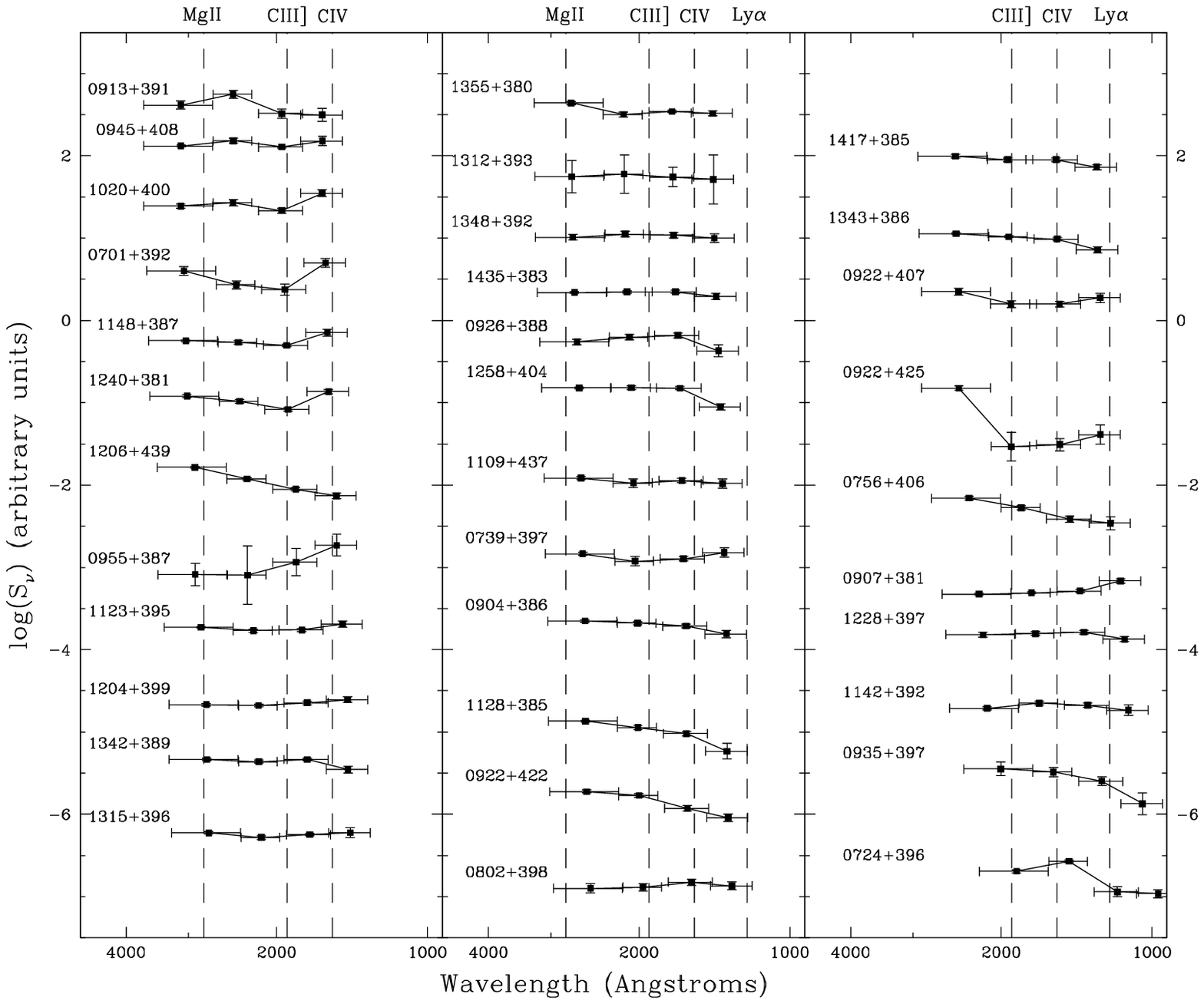}
\vskip -2truecm
\contcaption{}
\label{}
\end{figure*}

\begin{figure}
\epsfxsize=14cm
\vskip -1.2truecm
\epsffile{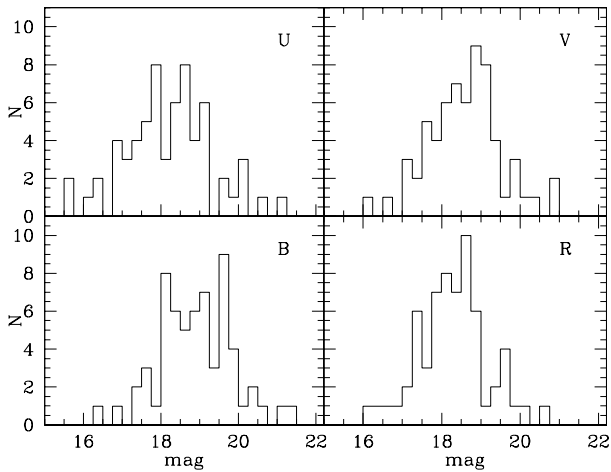}
\vskip -6.6truecm
\caption{Histograms of the $UBVR$ magnitudes for the 64 quasars with good
photometry}
\label{}
\end{figure}

\vfil\eject

\noindent
from the literature, which
can be used to estimate their average contribution to the broad-band
fluxes. The listed EWs include the values obtained by Baker \&
Hunstead (1995) for Molonglo radio quasars, which is a sample selected
at 408 MHz, with $S_{408}> 0.95$ Jy, and therefore appropiate 
for comparison with our sample (although it does not list
the EWs for the Fe{\sc ii} bumps). Also listed are the EW measurements for
radio-loud quasars given by Zheng et al. (1997, from an UV composite which
extends to 3000 \AA) and the EWs for optically-selected LBQS quasars
from Francis et al. (1991) and P.J. Green (1998). 
Francis et al.  (1991) mention that the height of the bumps
derived from their work would be decreased if the regions around 1700
\AA\ and 2650 \AA\ were taken as continuum, although at the cost of a
strongly curved continuum. Green (1998) uses as continuum the regions
2645--2700 \AA, 3020--3100 \AA\ and other wavelengths around 4430 and 
4750 \AA, and finds that the Fe{\sc ii}+Balmer bump in the range
3100--3800 is absent in his composite. The EW for Fe {\sc ii} 2400 
from Green (1998) is very similar to that measured by Zheng et al. (1997).

Baker \& Hunstead (1995) obtain qualitative estimates of the
contribution of Fe{\sc ii} bumps for different radio morphologies from the
height of the composite spectra relative to a power-law continuum
from $\sim$2000 to $\sim$5000 \AA. The authors conclude that the Fe{\sc ii}
bumps are absent in lobe-dominated and compact-steep-spectrum quasars,
but rather strong in core-dominated quasars, which resemble in
various line properties optically-selected quasars. The continuumm
of optically selected quasars is known to curve at around 3000 \AA\
(Francis et al. 1991, Natali et al. 1998), and the strength of the
bumps obtained by Baker \& Hunstead will be significantly decreased if
the underlying continuum was allowed to curve.
Boroson \& R.F. Green (1992) had previously found that flat-spectrum
quasars have stronger Fe{\sc ii} 4434--4684 \AA\ than steep-spectrum
ones, and that, as a whole, radio-loud quasars have fainter 
emission than radio-quiet ones. The latter result was also reported by
Cristiani \& Vio (1990); Fe{\sc ii} 2400 is the only Fe{\sc ii} 
bump revealed in their composite spectra for radio-loud
quasars, and the feature, almost absent, is less pronounced than in 
the composite for radio-quiet quasars. Zheng et al. (1997) derive
however a lower EW for Fe{\sc ii} 2400 for radio-quiet quasars (22
\AA\ versus 38 \AA), showing evidence of the uncertainties in the
contribution of these bumps.
 
Adopting as EWs for the C{\sc iv}, C{\sc iii}] and Mg{\sc ii} lines
the values from the Molonglo sample we infer the maximum contribution
to the broad-band fluxes for the C{\sc iv} line, amounting on average
to 25 per cent if the line is included in the $U$, $B$ or $V$
band. For the C{\sc iii}] line the average contribution is below 7 per
cent and for Mg{\sc ii} in the range 6--11 per cent. As EW for Fe{\sc
ii} $\sim2400$ \AA\ we adopted the value for the radio quasar sample
by Zheng et al. (1997). The EWs for C{\sc iv}, C{\sc iii}] and Mg{\sc
ii} by Zheng et al. (1997) are in fact very similar to those from the
Molonglo quasars, and the radio sample is also very similar to ours in
terms of optical luminosity. From the average EW of 38 \AA\ the
inferred contribution to the broad-band fluxes is 5--9 per cent.
The Fe{\sc ii}+Balmer $\sim$ 3400 feature lies outside the spectral region
covered by the composite by Zheng et al., and is absent in the
composites by Green (1998) and Cristiani \& Vio
(1990). Adopting half the EW from Francis et al. (1991),
i.e. around 38 \AA, its expected contribution would be below 6 per cent
when included in the $V$ or $R$ bands.

\begin{table}
\caption{Average equivalent widths of quasar emission lines}
\begin{tabular}{@{}lccccc}
Line       &Baker \&  &Zheng &Green&Francis&Adopted\\
           &Hunstead  &et al.&  &et al.& \\  
\hline
C{\sc iv}   &93&77 &68    & 37&93\\
C{\sc iii}] &23&17 &      & 20&23\\
Mg{\sc ii}  &58&50 &53    & 50&58\\
Fe{\sc ii} 2400   &  &38 &39    & 35&38\\
Fe{\sc ii} 3400$^*$  &  &   &absent& 76&38\\
Fe{\sc ii} 4500   &  &   &29    & 41&  \\
\hline
\end{tabular}
$^*$ Includes Balmer emission
\end{table}

\begin{figure}
\epsfxsize=15cm
\vskip -1.2truecm
\epsffile{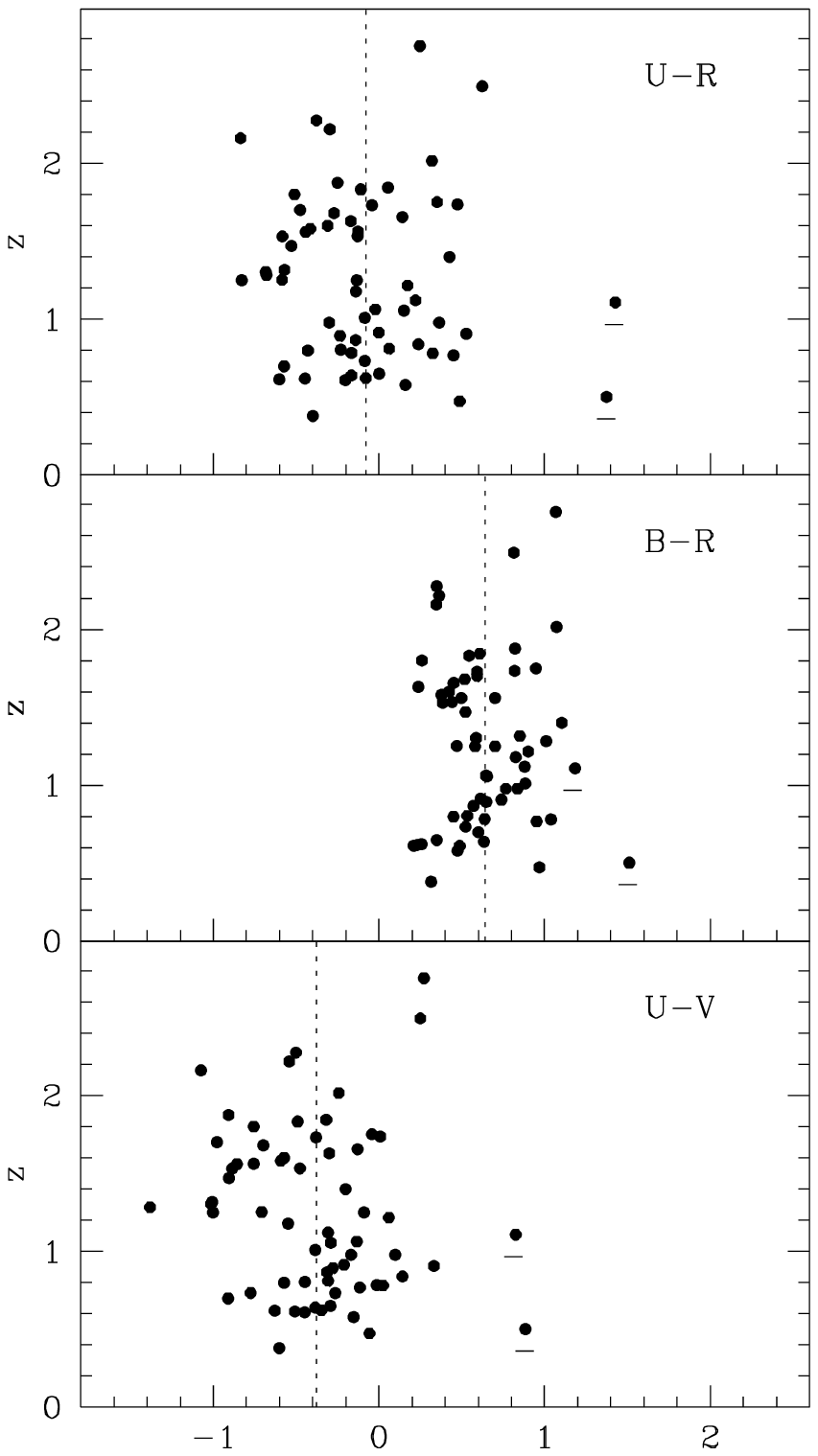}
\vskip -2.truecm
\caption{Distribution of the broad-band colours $U-R$, $B-R$ and $U-V$ as a
function of redshift. The underlined symbols correspond to the red quasars B3 
0918+381 and B3 1339+472 (see Section 6.1)}
\label{}
\end{figure}

\begin{table*}
\begin{minipage}{160mm}
\caption{Results of power-law and quadratic fits}
\begin{tabular}{@{}lcccclcccclccc}
B3 name & $\alpha$ & $\sigma(\alpha)$ &Model & &
B3 name & $\alpha$ & $\sigma(\alpha)$ &Model & &
B3 name & $\alpha$ & $\sigma(\alpha)$ &Model\\
\hline
0701+392 &~0.22& 0.25&q& & 0935+397 &-1.10& 0.41&p& &             1228+397 &-0.10& 0.13&p\\             
0704+384 &-0.81& 0.25&p& & 0937+391 &~0.37& 0.11&p& &		  1229+405 &-0.32& 0.17&q\\             
0724+396 &-0.49& 0.16&-& & 0945+408 &-0.04& 0.08&p& &		  1239+442B&-0.17& 0.08&p$\rightarrow$q\\       
0739+397B&-0.18& 0.14&p$\rightarrow$q& & 0951+408 &-0.27& 0.14&-& &	  1240+381 &-0.09& 0.10&-\\             
0740+380C&-0.41& 0.12&q& & 0953+398 &-0.56& 0.10&p& &		  1247+450A&-0.008&0.07&p\\             
0756+406 &-1.18& 0.18&p& & 1007+417 &~0.44& 0.05&-& &		  1256+392 &-1.11& 0.10&q\\             
0802+398 &~0.19& 0.25&p& & 1015+383 &~0.11& 0.09&p& &		  1258+404 &-0.29& 0.10&q\\             
0821+394 &-0.68& 0.15&-& & 1020+400 &~0.20& 0.15&-& &		  1315+396 &-0.09& 0.11&p\\             
0821+447 &-0.28& 0.16&-& & 1030+415 &-0.90& 0.13&p& &		  1339+472 &-1.73& 0.20&q\\             
0827+378 &-0.52& 0.07&q& & 1105+392 &-1.16& 0.20&p& &		  1341+392 &-1.06& 0.29&p\\             
0829+425 &-0.53& 0.17&p& & 1109+437 &-0.18& 0.14&p& &		  1342+389A&-0.13& 0.09&p\\             
0849+424 &-0.41& 0.29&p& & 1116+392 &-0.22& 0.11&q& &		  1343+386 &-0.43& 0.08&p\\             
0904+386 &-0.37& 0.11&p& & 1123+395 &-0.07& 0.09&p$\rightarrow$q& &	  1348+392 &~0.02& 0.19&p\\             
0907+381 &~0.36& 0.08&p& & 1128+385 &-0.80& 0.12&p& &		  1355+380 &-0.41& 0.11&p\\             
0910+392 &-0.40& 0.11&p& & 1141+400 &-0.83& 0.19&q& &		  1357+394B&-0.27& 0.10&p\\             
0913+391 &-0.52& 0.28&p& & 1142+392 &~0.12& 0.14&p$\rightarrow$q& &	  1416+400 &-1.14& 0.19&p\\             
0918+381 &-1.61& 0.28&p& & 1144+402 &-0.85& 0.11&p& &		  1417+385 &-0.36& 0.12&p\\             
0922+407 &-0.38& 0.22&p$\rightarrow$q& & 1148+387 &-0.14& 0.08&-& &	  1419+399 &~0.08& 0.11&-\\             
0922+422 &-1.02& 0.14&p& & 1148+477 &-0.25& 0.06&q& &		  1435+383 &-0.05& 0.01&p \\
0923+392 &-0.22& 0.06&-& & 1204+399 &~0.15& 0.10&p$\rightarrow$q& &\\
0926+388 &~0.12& 0.20&p& & 1206+439B&-1.28& 0.08&p& &\\
\hline
\end{tabular}
\end{minipage}
\end{table*}

\subsection{Optical broad-band colours of the quasars}

The distribution of the $U-R$, $B-R$ and $U-V$ colours as a
function of redshift is shown on Figure 3.  The broad-band colours do
not show a clear variation with redshift. If any, this would be a
blueing of the $U-V$ colour with redshift up to $z$ around 2.5. 
The mean colours derived for these quasars 
(indicated on the figure with dotted lines) are
$\langle U-R\rangle=-0.08$, $\langle B-R\rangle=0.64 $ and $\langle
U-V\rangle=-0.38$ with dispersions 0.45, 0.27, and 0.42, respectively, and 
correspond to observed spectral indices $\alpha_{\rm 
obs ~U-R}=-0.50$, $\alpha_{\rm obs ~B-R}=-0.39$ and $\alpha_{\rm 
obs ~U-V}=-0.75$ with dispersions 0.64, 0.53, and 0.96. From
the comparison of the $B-R$ and $U-V$ spectral indices a trend is
found in the sense that the SED is steeper at higher frequencies, in
agreement with the results found by Natali et al. (1998).

\section{Spectral energy distribution of the individual sources}

In this section we discuss the shape of the individual SEDs of the B3
quasars. The SEDs of the
61 sources with available photometry at the four bands were
fitted through $\chi^2$ minimization, using
a power-law model and a quadratic model. The
second model was chosen as a simple representation for the SEDs curved on
the log$S_\nu$--log$\nu$ plane.  A fit was accepted if the probability
$Q$ that the $\chi^2$ should exceed the particular measured value by chance
was higher than 1 per cent.
The best-fit spectral indices and their errors are listed in Table 4. 
Forty-one quasars have acceptable fits as power-laws. Thirty-nine of 
these have also acceptable fits as quadratics and in 6 cases the 
quadratic model gives a
significant improvement of the fits (higher than 85 per cent using an
$F$-test).  These six cases are labelled as p$\rightarrow$q on the last
column of Table 4. 
There are cases where the shape of the SED is clearly curved, but a
power-law model is acceptable due to large errors. Similarly, SEDs
which resemble power-laws to the eye and have small photometric
errors may not have acceptable fits for this model.  The two sources
with acceptable fits as power-laws but not as quadratics are B3
0849+424 and B3 1144+402, and both sources have poor power-law fits
($Q<0.02$). The SEDs of these sources could be contaminated 
by emission lines at the $U$ and $V$ bands.

Ten additional quasars without acceptable fits as power laws can be fitted
with quadratics. The remaining 10 sources do not have acceptable
fits with any of the models (indicated with an hyphen in
the last column of Table 4). For some of these sources the lack of good fits 
could be due to the presence of emission peaks in their SEDs related
to contamination by emission lines. We note for instance that 
contamination by C{\sc iv} 
could be related to the maxima in the $U$ band for B3 1020+400, B3 1148+387 
and B3 1240+381, and in the $V$ band for B3 0724+396. 

\begin{figure}
\vskip -2.3truecm
\epsfxsize=10cm
\epsffile{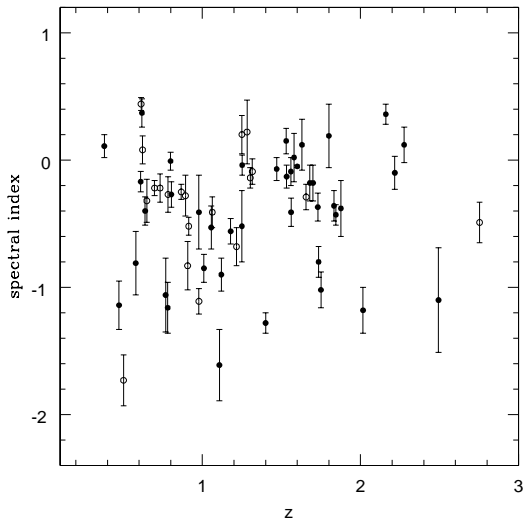}
\vskip -0.8truecm
\caption{Spectral index versus redshift for the 61 quasars with good 
photometry and available data in the four bands. 
Filled circles correspond to power-law fits formally acceptable}
\label{}
\end{figure}
\begin{figure}
\vskip -2.3truecm
\epsfxsize=10cm
\epsffile{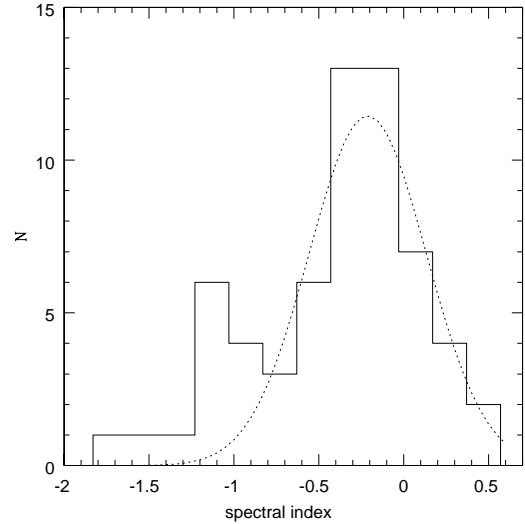}
\vskip -0.8truecm
\caption{Spectral index distribution for the quasars shown on Fig. 4. 
The dotted curve shows the gaussian fit for the quasars with $\alpha>-0.9$}
\vskip -0.1truecm
\label{}
\end{figure}

Figure 4 shows the distribution of spectral indices versus redshift
for the 61 sources. The slopes were obtained from fixed observed
wavelengths, therefore they correspond to different rest-frames 
(from 2600--5000 for $z=0.4$ to 1000--1900 for $z=2.8$).  The
slopes show a wide range, with values from 0.4 to --1.7, regardless
that the fits are formally acceptable or not. The mean and the
dispersion for the total sample are --0.39 and 0.38 respectively
(--0.41 and 0.40 for the sources with formally acceptable power-law fits). 
The comparison of these values with those obtained from two-band colours
(Sect. 4.2) show the best agreement for $B-R$, with $\alpha_{\rm obs
~B-R}=-0.39$ and standard deviation 0.53. The reason for this is that
the $B$ and $R$-band errors are low, compared to those for the
$U$-band, and the errors are weighted for the power-law fits.

Although a mean spectral index was obtained from the sample, the
distribution of slopes is asymmetric, showing a tail to steep indices
(see Fig. 4 and the histogram on Figure 5). Considering the total
sample of 61 sources, the distribution of spectral indices for
$\alpha>-0.9$ is well represented by a gaussian 
with a mean of --0.21 and a dispersion of 0.34.

\section{Spectral energy distribution from the composite SEDs}

\subsection{Normalized composite SEDs}

The shape of the SED of the quasars can be analysed through a
``composite spectrum'', in which the individual SEDs are merged in the
rest-frame, adopting a specific criterium for the scaling of the
fluxes. The coverage of the optical photometry 
and the range of
redshifts of the quasars allows to study their spectrum in the
range $\sim$$1300-4500$ \AA. We chose to normalize the
individual SEDs to have the same flux density at a fixed wavelength
$\lambda_{\rm n}$ {\it within} the observed range, and the flux
density at $\lambda_{\rm n}$ was obtained by linear interpolation
between the two nearest data points.  With this procedure we could not
have an appropriate normalization for the whole sample (the coverage for 
the lowest redshift quasars is $2500-5000$ \AA\ and for the highest
redshift ones $1000-2000$ \AA), therefore we obtained composite SEDs
for different normalization wavelengths.
The selected $\lambda_{\rm n}$ were 3800, 3500, 3200,
2400, 2200 and 2000 \AA\ and the corresponding normalized SEDs are
shown on Figure 6. The use of a broad range of values for
$\lambda_{\rm n}$ is appropriate to analyse the possible 
dependence of the SED shape with $\lambda_{\rm n}$. The low separation
between the $\lambda_{\rm n}$ values, typically around 300 \AA, allows
for a large overlap between the composites and for their comparison in
a continuous way. A normalization around 2800 \AA\ was not considered,
to avoid the Mg{\sc ii} emission line. The fluxes at the normalization 
$\lambda_{\rm n}$=2000 could be contaminated by the weaker 
C{\sc iii}]$\lambda$1909 line, and those at the normalizations 
2400, 3200 and 3500 by the emission bumps at 2250--2650 and
3100--3800.  
However, we have seen in Sect. 4 that the expected
contributions of these features are weak; typically below 10 per
cent. A 10 per cent contribution corresponds to
a vertical shift in log$S_\nu$ (Figure 6) of 0.04, which is clearly
lower than the dispersion of the spectral points in the composites. 

\begin{figure*}
\epsfxsize=16cm
\epsfysize=16cm
\vskip -0.8truecm
\epsffile{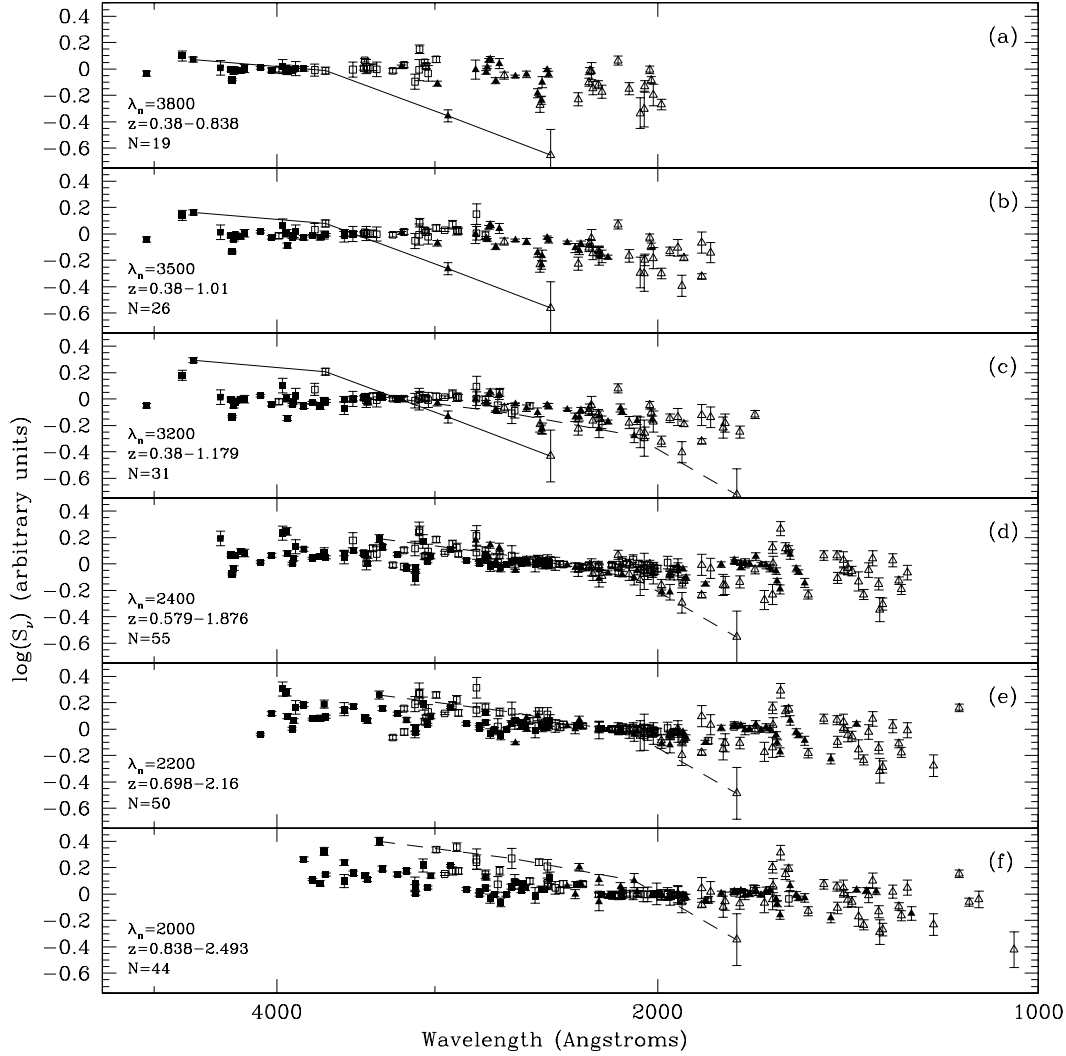}
\vskip -1.0truecm
\caption{Composite SED for different normalization wavelengths. Spectral
points from different optical bands are shown with different
symbols. Filled squares, empty squares, filled triangles and empty
triangles correspond respectively to $R$,$V$,$B$ and $U$-band data.  The
spectral points of the red quasars B3 0918+381 and B3 1339+472 appear
connected on the figure to distinguish their individual SEDs.}
\label{}
\end{figure*}

\begin{table*}
\begin{minipage}{160mm}
\caption{Power-law fits in the intervals 4500--3000 \AA~ and 
2600--1700 \AA~ for quasars in the redshift range 0.38--2.49}
\begin{tabular}{@{}clllllllllllr}
$\lambda_{\rm n}$& $\alpha_{\rm blue}$ & $\sigma$&
N&$z$ range&~~~&$\alpha_{\rm UV}$ & $\sigma$&N&$z$ range&&
$\alpha_{\rm blue}-\alpha_{\rm UV}$ & $\sigma$\\
\hline
$3800$&~0.20     &0.12 & 34& 0.38-0.84&~~~&          &    &    &          &\\
$3500$&~0.23     &0.11 & 41& 0.38-1.01&~~~&-0.73$^*$ &0.35& 38 & 0.47-1.01&& 0.96$^*$& 0.37 \\
$3200$&~0.21     &0.12 & 45& 0.38-1.18&~~~&-0.82     &0.25& 47 & 0.47-1.18&& 1.03    & 0.28 \\
$2400$&-0.08    &0.23 & 50& 0.57-1.32&~~~&-0.61     &0.11& 96 & 0.57-1.88&& 0.53    & 0.25 \\
$2200$&-0.25$^*$&0.35 & 36& 0.69-1.32&~~~&-0.57     &0.10& 93 & 0.69-2.16&& 0.32$^*$& 0.36 \\
$2000$&         &     &   &          &~~~&-0.66     &0.10& 80 & 0.83-2.49&&\\
\hline
\end{tabular}

\end{minipage}
\end{table*}

Some of the spectral points clearly deviating in the composite SEDs
correspond to the two reddest quasars, B3 0918+381 and B3 1339+472
($\alpha \leq -1.6$). The extreme red colours of these two sources are
uncommon in the sample, as can be seen from Fig. 3, where the symbols
for the two sources appear underlined. Both sources show curved
spectra in the log$S_\nu$--log$\lambda$ plane.  B3 0918+381 has an
acceptable fit for the power-law model, but this is due to the large
magnitude error at the $U$-band.  On Fig. 6 the normalized SEDs of
these sources are indicated over the remaining spectral points,
showing the notorious discrepancy relative to the average composite
SEDs. These two sources with peculiar SEDs will not be considered for
the discussion in this and the next section.

The overall shape of the composite spectra is found to be very uniform
for the different normalizations. The dispersion of the spectral
points is artificially reduced near the wavelength
$\lambda_{\rm n}$. The spectral points appear to trace predominantly
the continuum; only the C{\sc iv}$\lambda$1549 line appears prominent
on the composites, as an ``emission feature''.  No other
emission bumps/lines are apparent from the composites, 
in agreement with the low contributions expected (Sect 4.1). 
The C{\sc iv} feature
is revealed in the three composites covering its wavelength and arises
from the $U$-band data of various quasars, most of which were
mentioned in Sect. 5 to have likely contamination by this line.  The
emission feature appears in the broad-band composite displaced to the
red, peaking at around 1610 \AA. The displacement is around five times
lower than the rest-frame width over which the line is detected.

A general trend on Fig. 6, apparent especially from panels (b) to (d),
is a steepening of the spectrum from large to short wavelengths,
ocurring at around 3000 \AA. Above this wavelength the SED appears to
be rather flat. This trend is less evident on panel (a), probably
because of the small number of spectral points at $\lambda<3000$
\AA. The trend is weak in panel (e) and practically disappears in
panel (f), but here the number of points with $\lambda>3000$
\AA\ is small. Panels (d) and (e) appear to show that the SED flattens
again below 2000 \AA. This trend practically dissapears in panel (f),
in spite of the large number of points, thus this 
flattening is not so clear in principle as the steepening at 3000 \AA.

\subsection{Power-law fits}

We have obtained power-law fits of the composite SEDs in the regions
above and below the 3000 \AA\ break. The selected ranges were
4500--3000 \AA\ (referred to as ``blue'') and 2600--1700 \AA\
(referred to as ``UV''), which roughly correspond to the regions
limited by H$\beta$ and Mg{\sc ii}$\lambda$2798 and Mg{\sc
ii}$\lambda$2798 and C{\sc iv}$\lambda$1549. These spectral points
were taken as representing the continuum, since the contamination by
emission lines/bumps in this range is expected to be very weak. The
results of the fits, obtained by least-squares minimization, are
listed in Table 5, including for each composite the slopes and their
errors, the number of spectral points used and the redshift range of
the quasars. Slopes with an asterisk correspond to lower quality fits,
since the spectral points do not cover the whole wavelength
range. Figure 7 shows the spectral index values as a function of
$\lambda_{\rm n}$, illustrating its variation between the different
composites.  The four composite SEDs for which power-laws were fitted
in both ranges show a steepening from low to high frequency. The
differences $\alpha_{\rm blue}-\alpha_{\rm UV}$ for these composites
and their errors are listed in Table 5. For the following discussion
only the good quality fits will be considered.

Concerning the 4500$-$3000 \AA\ range, the spectral indices for all
the normalizations are consistent within their errors. The first three
normalizations, with $\lambda_{\rm n}$=3800, 3500 and 3200, show a
better agreement with each other.  Averaging the spectral indices for
the fits for $\lambda_{\rm n}$=3800, 3200 and 2400 (the fit for 3500
\AA\ has a large overlap with those for 3200 and 3800 \AA) we obtain
$\langle\alpha_{\rm blue} \rangle = 0.11 \pm 0.16$.

The high frequency fits have again best-fit slopes consistent with
each other within the errors. For the high frequency normalizations at
$\lambda_{\rm n}$=2400, 2200 and 2000, the spectral indices are very
similar, although the range of spectral points used do not show a wide
overlap. The spectral index for the low frequency normalization at
$\lambda_{\rm n}$=3200 is steeper.  Averaging the slopes obtained for
$\lambda_{\rm n}$=3200, 2400, 2200 and 2000 we find $\langle
\alpha_{\rm UV} \rangle = -0.66 \pm 0.15$.  
Considering the average measured values of $\alpha_{\rm blue}$ and
$\alpha_{\rm UV}$ and their errors we find a steepening towards high
frequencies $\langle \alpha_{\rm blue} \rangle - \langle \alpha_{\rm
UV} \rangle = 0.77 \pm 0.22$. 

Although the expected contribution of emission lines/bumps in the
ranges used for the power law fits is low, it is interesting to
analyse the possibility that the slope change is artificially produced
by contamination due to the Fe{\sc ii} bumps at 3100--3400 and
2250--2650, since these features would enhance the emission used for the 
power-law fits (4500--3000 and 2600--1700 \AA) in the regions next to
the break, producing a slope change in the observed sense. 
However, a power law fit for the whole range from
1700 to 4500 \AA\ for the composites (b) to (d), where the break is
more obviously detected, yields a clear excess emission only in the
region 2650--3200 \AA\, i.e. between the two bumps. In fact some
excess there could be due to Mg{\sc ii}, but if the Fe{\sc ii} bumps
were responsible for the slope change, the excess emission should
extend over the Fe{\sc ii} ranges 2250--2650 and 3100--3400.
Therefore, although some of the broad-band spectral points are
expected to include contamination due to emission lines and/or bumps,
the break in the overall SED detected at $\sim$ 3000 \AA\ is most
likely related to an intrinsic change in the continuum of the quasars.
A steepening of the quasar's continuum at 3000 \AA\ was reported by
Natali et al. (1998) from spectra of optically selected quasars, and
is also found in the composite spectrum of optically selected quasars
by Francis et al. (1991). The slopes obtained by Natali et al. (1998)
were $\langle \alpha_{\rm blue} \rangle \simeq 0.15$ for the range
2950--5500 \AA, and $\langle
\alpha_{\rm UV} \rangle \simeq -0.65$ for the range 1400--3200 \AA, in
good agreement with our values.

\begin{figure}
\epsfxsize=9.5cm
\vskip -0.3truecm
\epsffile{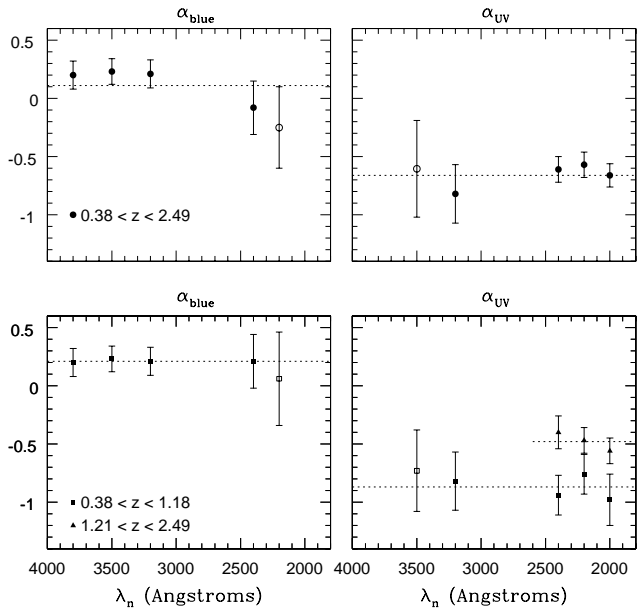}
\vskip -1.truecm
\caption{Spectral indices as a function of $\lambda_{\rm n}$. 
Empty symbols correspond to poor fits. Top panels correspond to the 
results in Table 5, for quasars of all redshifts, and the
bottom panels to the results in Table 6, separating low and high
redshift quasars}
\label{}
\end{figure}

\subsection{Redshift dependence of the blue/UV continuum shape}

A striking characteristic of the best-fit slopes on Table 5 and Fig. 7
(top panels) is how they keep roughly constant for the normalizations
at either side of the break around 3000 \AA, suffering the largest
variation when the normalization moves from one side of the break to the 
other. In particular, the change in $\alpha_{\rm UV}$ from
$\lambda_{\rm n}$=3200 to lower values of $\lambda_{\rm n}$ is
consistent with the description outlined on Section 6.1 of a possible
flattening at around 2000 \AA\ for the composites with $\lambda_{\rm
n}$=2400 and 2200 \AA. It is important to note that the change in
normalizations from $\lambda_{\rm n} \geq 3200$ \AA\ to $\lambda_{\rm
n} \leq 2400$ \AA\ in our study implies a substantial {\it change in
the redshifts} of the quasars whose spectral points enter the
fits. For instance, whereas that the high frequency fit for
$\lambda_{\rm n}$=3200 comprises 47 spectral points with
$0.47<z<1.18$, the next composite, with $\lambda_{\rm n}$=2400,
includes all but one of these points plus 50 additional spectral
points from $1.21<z<1.88$. For the final composite, with $\lambda_{\rm
n}$=2000, only 23 out of the 80 spectral points used for the fit are
in common with the fit for $\lambda_{\rm n}$=3200.  A similar
variation of the spectral points/redshifts with $\lambda_{\rm n}$
occurs for the low frequency fits, although in this case the low
redshift limit for $\lambda_{\rm n}$=3800, 3500 and 3200 is similar
and the variations of spectral points/redshifts are lower for all the
normalizations (see Table 5).  It is therefore interesting to analyse
whether a dependence of the shape of the continuum spectrum with
redshift is present.

\begin{figure*}
\epsfxsize=15cm
\epsfysize=15cm
\vskip -1truecm
\epsffile{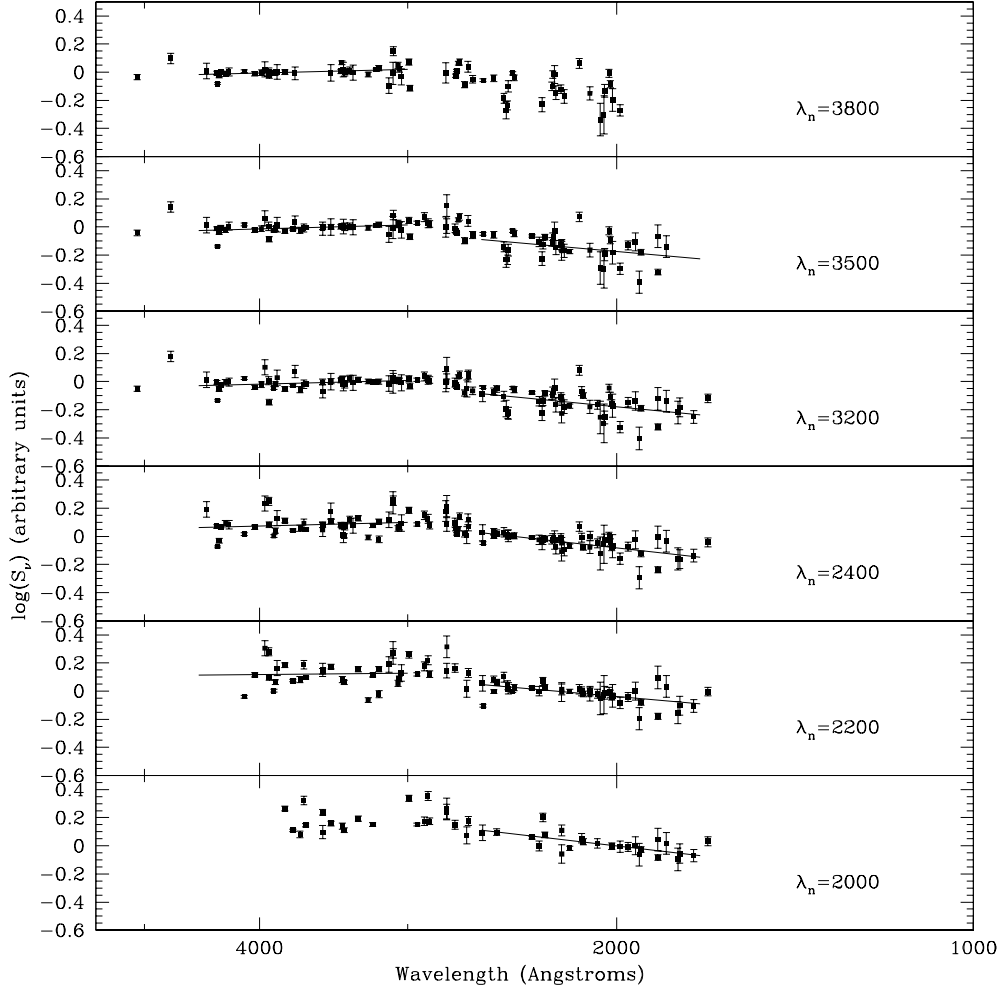}
\vskip -0.8truecm
\caption{ Same as Fig. 6 for quasars with 
$z<1.2$. The straight lines show the 
best-fit power laws in the regions 4500--3000 \AA\ and 2600--1700 \AA}
\label{}
\end{figure*}

\begin{figure*}
\epsfxsize=15cm
\epsfysize=15cm
\vskip -1truecm
\epsffile{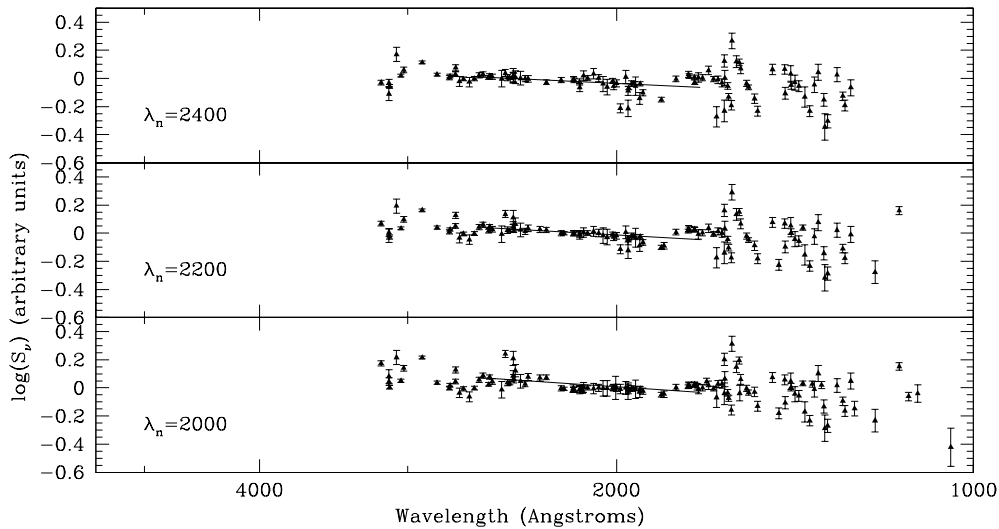}
\vskip -6.5truecm
\caption{ Same as Fig. 6 using only spectral points from quasars with 
$z>1.2$}
\label{}
\end{figure*}
 
From the results of the fits (Table 5, Fig. 7, and see also Fig. 6)
we decided to separate the quasar sample in a low redshift
bin with $z<1.2$ and a high redshift bin with $z>1.2$, and
perform the power-law fitting for the two subsamples. 
With the selected limit, all the spectral points used for
the normalizations with $\lambda_{\rm n} > 3000$ \AA\ correspond to
the low redshift bin. The composites with $\lambda_{\rm n} \leq 2400$
\AA\ include spectral points from quasars in the two redshift bins.
Figures 8 and 9 show the composite SEDs for the two quasar subsamples.
The results of the power-law fits are given in Table 6 and the variation of 
the slopes as a function of $\lambda_{\rm n}$
is shown on Fig. 7.

\begin{table*}
\begin{minipage}{160mm}
\caption{Power-law fits for quasars in the redshift range 0.38--1.18 (upper
part) and 1.21--2.49 (bottom part)}
\begin{tabular}{@{}clllllllllllr}
$\lambda_{\rm n}$& $\alpha_{\rm blue}$ & $\sigma$& N&
$z$ range&~~~&$\alpha_{\rm UV}$ & $\sigma$&N&$z$ range&& 
$\alpha_{\rm blue}-\alpha_{\rm UV}$ & $\sigma$\\ 
\hline
3800&0.20 &0.12 &34&0.38-0.84&~~~& & & & & &\\ 
3500&0.23 &0.11&41&0.38-1.01&~~~&-0.73$^*$&0.35&38&0.47-1.01&&0.96$^*$&0.37\\ 
3200&0.21&0.12 &45&0.38-1.18&~~~&-0.82 &0.25&47&0.47-1.18&&1.03 &0.28\\ 
2400&0.21&0.23 &43&0.57-1.18&~~~&-0.94 &0.17&48&0.57-1.18&&1.15 &0.29\\
2200&0.06$^*$&0.40&29&0.69-1.18&~~~&-0.76&0.17&41&0.69-1.18&&0.82$^*$&0.43\\ 
2000& & & & &~~~&-0.98&0.22&24&0.83-1.18&& &\\ & & & & &~~~& & & & & &\\ 
2400& & & & &~~~&-0.40&0.14&48&1.21-1.88&& &\\ 
2200& & & & &~~~&-0.47 &0.11&52&1.21-2.16&& &\\
2000& & & & &~~~&-0.56 &0.11&56&1.21-2.49&& &\\ 
\hline
\end{tabular}

%$^*$For these fits the spectral points do not cover the whole selected 
%wavelength range

\end{minipage}
\end{table*}

Concerning the low redshift quasars, we find now a better agreement
between the slopes of the low-frequency fits obtained for the
normalizations on either side of the break. 
Averaging the values for the normalizations with
$\lambda_{\rm n}=3800$, 3200 and 2400 we find $\langle \alpha_{\rm
blue} \rangle = 0.21 \pm 0.16$.  For the high frequency fits there is
also a better agreement in $\alpha_{\rm UV}$ between the
normalizations at either side of the break. The agreement is due to
the steepening of $\alpha_{\rm UV}$ for the normalizations with
$\lambda_{\rm n} \leq 2400$ \AA\ when high redshift quasars have been
excluded. The average value of $\alpha_{\rm UV}$ is now $\langle
\alpha_{\rm UV} \rangle = -0.87 \pm 0.20$. We find now a 
larger change from $\alpha_{\rm blue}$ to
$\alpha_{\rm UV}$, with 
$\langle \alpha_{\rm blue} \rangle -
\langle \alpha_{\rm UV} \rangle = 1.08\pm0.26$. The larger steepening
is evident from the comparison of the top and bottom panels 
of Fig.  7. The good agreement in the slopes obtained for the different 
normalizations indicates that the selection of $\lambda_{\rm n}$ does not 
affect the derived slopes. 

For the high redshift quasars only the high frequency region can be
studied (see Fig. 9). The average spectral index from the three
normalizations is $\langle \alpha_{\rm UV} \rangle = -0.48 \pm 0.12$.
Comparing the values of $\alpha_{\rm UV}$ for low and high redshift
quasars we find a difference $\Delta\langle \alpha_{\rm UV}\rangle =
0.39\pm0.23$.  This difference although weak, is larger than the
error, and is clearly evident from the results on Table 6 and Fig. 7, and 
the comparison of figs. 8 and 9.
The flattening below around 2000 \AA\ noted on Sect. 6.1 for the high
frequency normalizations is related to this hardening of $\alpha_{\rm
UV}$ for high redshift quasars.

Our result of a dependence of the spectral index with redshift, in the
sense that quasars at $z>1.2$ tend to be flatter, i.e. harder, than 
low redshift ones, in the range 2600--1700 \AA, is similar to that found by 
O'Brien et al. (1988) for an $IUE$-selected sample, with slopes in the region
1900--1215 \AA\ ranging from --0.87 for $z<1.1$ to --0.46 for $z>1.1$. We 
note the remarkable agreement between these values and those obtained for our 
sample. This trend with redshift was not confirmed by Natali et al. (1998) 
for their sample of optically selected quasars, although their limits for the 
detection of a variation were around 0.4, which is comparable to the measured 
change in the slope found in O'Brien et al. and in this work.

A possible concern on our results is whether the 20 per cent
incompleteness of the sample could produce a bias against the
inclusion of red quasars at high redshifts. However, the distribution
of $R$ magnitudes in the range $z$=0.5--2.5 is roughly constant, and
does not suggest that the fraction of missed quasars ($R>20$) should
be higher at the higher redshifts. On the other hand, 5 good B3-VLA 
quasar candidates excluded from the B3-VLA Quasar Sample were present 
on the blue POSS-I plate but not on the red POSS plate. Since the 
magnitude limit for POSS blue is around 20.5--21, these quasars 
have $B-R$ colours similar or bluer than the average value 
for the quasars in the sample ($B-R$= 0.64, Sect. 4.2). Although this number 
of quasars is small, it illustrates the presence of blue or normal 
colours among the missed quasars. Since the incompleteness does not appear to 
be biased towards the exclusion of red quasars at high redshifts, 
the $\alpha_{\rm UV}-z$ trend is likely a real effect, not originated 
by incompleteness.

\subsection{Relation of the blue/UV continuum shape to radio power}

Figure 10 presents the $P_{408}-z$ diagram for the 73 quasars observed
for this work.  In agreement with the expectations
for flux-limited samples, high redshift quasars tend to have
higher radio powers than low redshift ones, although the B3-VLA
Quasar Sample is not strictly a flux-limited one (see Sect. 2). Using
Spearman's correlation coefficient $r_s$, we find for the trend 
$r_s$=0.52, with a significance level $P>99.999$. In order to check whether 
the $\alpha_{\rm UV}-z$ trend could arise from an
intrinsic  $\alpha_{\rm UV}-P_{408}$ correlation and the $P_{408}-z$ trend, we
performed an analysis similar to the one presented in sections 6.2 and 6.3,
obtaining separated power-law fits for a ``high-radio-power subsample'' and
a ``low-radio-power subsample''. Instead of considering several composites, 
the dependence on radio power was analysed
using only the composite with $\lambda_{\rm n}$=2400. 
This composite includes the largest number of quasars, and the
redshift distribution comprises low as well as high redshift ones,
with a median $z$ of 1.15, similar to that of the whole
sample.

The composite with $\lambda_{\rm n}$=2400 includes 54 quasars
(excluding B3 0918+381).  The median of $P_{408}$ is 
$10^{34.85}$ erg s$^{-1}$ Hz$^{-1}$ and this value was used as
the division limit between low and high radio power. The
low-power quasars have a mean redshift of 1.00 and a mean radio power 
$10^{34.48}$ erg s$^{-1}$ Hz$^{-1}$, and the same parameters for
the high-power quasars are 1.35 and $10^{35.19}$ erg s$^{-1}$
Hz$^{-1}$. The results of the derived spectral indices $\alpha_{\rm
UV}$ for the two subsamples are listed on Table
7. The table includes for comparison the $\alpha_{\rm UV}$ values 
obtained for all the quasars and for the subsamples separated by
redshift, for the same composite.  The third column of the table gives
$\Delta\alpha_{\rm UV}$ and its error for each couple of subsamples
with a varying parameter ($z$ or $P_{408}$), and the last
column gives the probability that the difference in slope is
statistically significant.

\begin{figure}
\epsfxsize=8cm
\vskip -1.0truecm
\epsffile{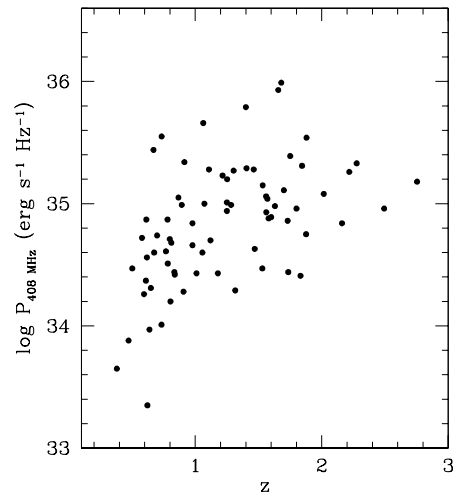}
\caption{$P_{\rm 408 ~MHz}$ versus redshift for the 73 B3-VLA quasars 
observed
for this work}
\label{}
\end{figure}

\begin{table}
%\begin{minipage}{160mm}
\caption{$\alpha_{\rm UV}$ for the composite
with $\lambda_{\rm n}$=2400 and different subsamples, separated by 
redshift, 
radio power and blue/UV luminosity}
\begin{tabular}{@{}lccc}
                &$\alpha_{\rm UV}$&$\Delta(\alpha_{\rm UV})$&Significance of\\
                &               &                       &the slope change\\
\hline
All $z$         &$-0.61 \pm0.11$&             &  \\
                &               &             &  \\
$z<1.2$         &$-0.94 \pm0.17$&$0.54\pm0.22$&98.84\%  \\
$z>1.2$         &$-0.40 \pm0.14$&             &  \\
                &               &             &  \\
$P_{408}<10^{34.85}$ &$-0.71 \pm0.18$&$0.20\pm0.23$&27\%  \\
$P_{408}>10^{34.85}$ &$-0.51 \pm0.14$&             &  \\
                &               &             &  \\
$L_{2400}<10^{30.67}$&$-0.76\pm 0.18$&$0.30\pm0.22$&89\%  \\    
$L_{2400}>10^{30.67}$&$-0.46 \pm0.13$&             &  \\
\hline
\end{tabular}
\end{table}

The largest difference in $\alpha_{\rm UV}$ and the
highest significance is found for the subsamples
separated by $z$, with $\Delta\alpha_{\rm UV}=0.54 \pm 0.22$, and
98.84 per cent significance ($F$-test). For the subsamples
separated by $P_{408}$ the difference in $\alpha_{\rm UV}$ is lower
than the errors. The trend
$\alpha_{\rm UV}-z$ is therefore unlikely to be a secondary correlation from an
intrinsic $\alpha_{\rm UV}-P_{408}$ correlation in conjunction with
the $P_{408}-z$ trend. Moreover, the last trend is less significant
for the group of 54 quasars in the $\lambda_{\rm n}$=2400 composite 
($r_s$=0.44 and $P$=99.906). This weak trend explains in fact
the reverse result, i.e. that the $\alpha_{\rm UV}-z$ trend in
combination with the $P_{408}-z$ trend does not cause a $\alpha_{\rm
UV}-P_{408}$ trend.

\subsection{Relation of the blue/UV continuum shape to blue/UV luminosity}

The possible relation between $\alpha_{\rm UV}$ and the blue/UV
luminosity was also investigated.  In order to construct each
composite SED we had to interpolate the flux density at the
normalization wavelength $\lambda_{n}$, and therefore we could obtain
the monocromatic luminosity at $\lambda_{\rm n}$ straightforward. The
possible trend $L_{\rm blue/UV}-\alpha_{\rm UV}$ was analysed in the same 
way as for the $P_{408}-\alpha_{\rm UV}$ trend, measuring $\alpha_{\rm UV}$ in
the composite with $\lambda_{\rm n}=2400$ for two subsamples
separated by $L_{2400}$. The separation value was set at
$L_{2400}=10^{30.67}$ erg s$^{-1}$ Hz$^{-1}$, which is the median
value for the quasars in the composite. The values of
$\alpha_{\rm UV}$ (Table 7) indicate a weak flattening with
luminosity, only slighly larger than the errors, but significant from
the $F$-test, with $P$=89 per cent. The significance is lower than
that found for the $\alpha_{\rm UV}-z$ trend.

Figure 11 shows $L_{2400}$ versus $z$ for the quasars in the same
composite. A clear correlation is found, in the sense that high
redshift quasars are more luminous, similar to the trend with radio
power. A Spearman correlation gives $r_s$=0.52 and significance 99.994
per cent. Therefore the three parameters $\alpha_{\rm UV}$, $z$ and
$L_{2400}$ are correlated with each other, being in principle
impossible to determine which are primary and which are secondary
correlations.  The quasars with luminosities below the limit of
$10^{30.67}$ erg s$^{-1}$ Hz$^{-1}$ have a mean redshift and
luminosity of 1.01 and $10^{30.27}$ erg s$^{-1}$ Hz$^{-1}$, and the
same parameters for the luminous quasars are 1.33 and $10^{31.02}$ erg
s$^{-1}$ Hz$^{-1}$.

\begin{figure}
\epsfxsize=8cm
\vskip -1.0truecm
\epsffile{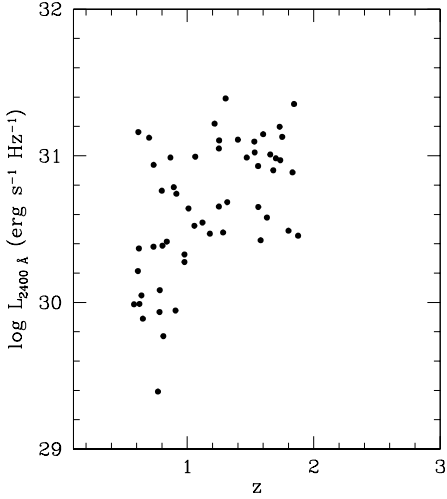}
\caption{$L_{\rm 2400 ~\AA}$ versus $z$ for the 54 quasars
with measured (interpolated) flux density at 2400 \AA}
\label{}
\end{figure}

\begin{figure*}
\epsfxsize=16cm
\epsfysize=16cm
\vskip -1.5truecm
\epsffile{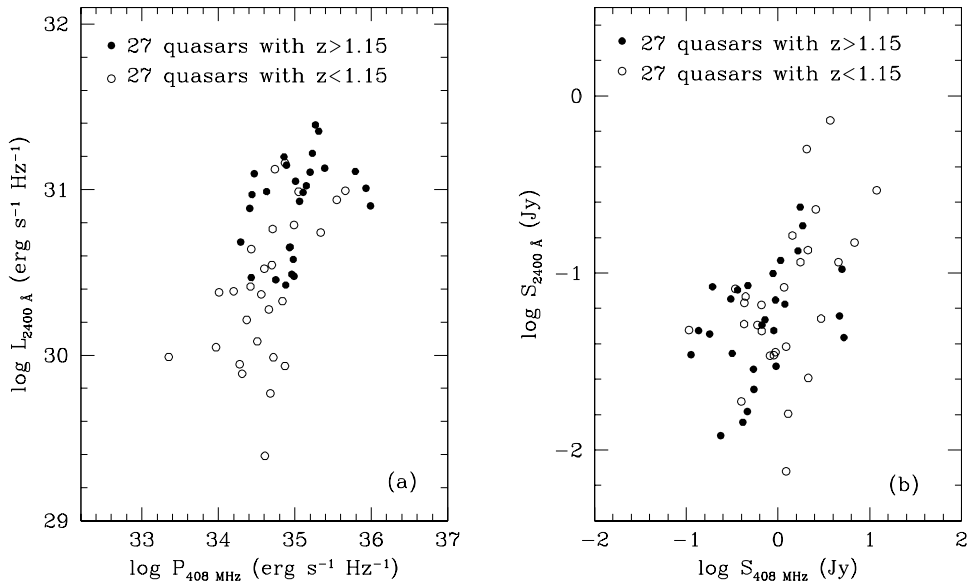}
\vskip -6.4truecm
\caption{Panel (a): $L_{2400}$ versus $P_{408}$ for 54 B3-VLA quasars in this
 work. Panel (b): $S_{2400}$ versus $S_{408}$ flux densities for the same
quasars. For the radio fluxes we used the $k$-correction 
$C=-(1+\alpha)$ log$(1+z)$, where $\alpha$ is the radio spectral index between
408 and 1460 MHz given by Vigotti et al. (1989). For the flux densities at 
2400 \AA\ the $k$-correction adopted was $C=-$log$(1+z)$}
\label{}
\end{figure*}

Wandel \& Petrosian (1988) obtained predictions of the expected UV
slope versus UV luminosity for various accretion disc models, spanning
a range of values for the black hole mass, accretion rate and
viscosity parameter. They found that the effect of a flattening of the
spectrum for the more luminous quasars (or at higher redshifts),
reported by O'Brien et al. (1988), could be explained as evolution
along curves of constant black hole mass, with accretion rate
decreasing with time. The emitted spectrum for these models under the
assumption of constant black hole mass is less luminous and more steep
in the UV as the accretion rate decreases, and a decrease in accretion
rate with time could explain the observed $L_{\rm UV}-z$ and
$\alpha_{\rm UV}-z$ trends.  Our data show the same trend of a
flattening of $\alpha_{\rm UV}$ with $z$ and $L_{\rm UV}$, as well as
a $L_{\rm UV}-z$ correlation, therefore the same interpretation could
apply.  The diagrams $\alpha_{\rm UV}-L_{\rm UV}$ presented by Wandel
\& Petrosian (1988) correspond to a slope in the range $1500-1000$
\AA\ and the luminosity at 1450 \AA. We have represented in these
diagrams the average slope for low-redshift and high-redshift quasars
(separation at $z=1.2$), using the composite with $\lambda_n=2400$,
and the 2600--1700 \AA\ slopes listed in Table 7. The luminosities
were obtained transforming the average values at 2400 \AA\ to 1450
\AA, adopting the measured slopes.  Although we use a slope 
different than that of the models, at least for high redshift quasars the
slope does not apperar to change between the two ranges (Fig. 9). A
straight line connecting the location of the two subsamples on the
$\alpha_{\rm UV}-L_{\rm UV}$ plane would run approximately parallel to
the curves of constant black hole mass, 
with $\dot{m}$ increasing in the same
direction as the redshift. Therefore our results appear to be consistent with
the interpretation outlined above, and the comparison with the models
yields a roughly constant black hole mass, in the range $M \simeq
10^{8.2-8.5} M_\odot$, and accretion rates (in units of the Eddington
value) ranging from $\dot{m} \simeq$ 1 for the quasars with
$z=0.6-1.2$ to $\dot{m} \simeq$ 10 for the quasars with $z=1.2-1.9$.

\section{The correlation between blue/UV luminosity and radio power}
 
Figure 12(a) shows $L_{2400}$ versus $P_{408}$ for the 54 quasars of
the $\lambda_{\rm n}$=2400 composite. $L_{2400}$ and $P_{\rm 408
~MHz}$ are strongly correlated, with $r_s$=0.62 and $P>99.9999$ per
cent. The optical completeness of the sample is high and we do not
expect important selection effects in the optical that could explain
the correlation.
The correlation could be in principle artificially induced by
independent evolution of both parameters with redshift. In fact the
quasars with high blue/UV luminosity tend to have large redshifts and those
weaker tend to have low redshifts, and a similar behaviour occurs for
the radio power. However, this alternative is ruled out by the
redshift distribution on Figure 12(a), showing that the half subsample with 
lower redshifts shows a similar $L_{2400}-P_{408}$ trend as the whole sample
(with unrestricted $z$). The correlation is confirmed in the flux-flux plane 
on Figure 12(b), with $r_s=0.47$ and $P=99.966$ over 
two decades in both radio and blue/UV flux. 

The most likely origin for the radio-optical correlation is therefore
a direct link between the blue/UV emission of the quasar nucleus, 
related to the accretion process, and the radio synchrotron
emission at 408 MHz. A similar correlation was reported by Serjeant et
al. (1998) on the basis of complete samples of radio-steep-spectrum
quasars (from Molonglo, 3CR and the Bright Quasar Survey), in the
redshift range 0.3--3.
A linear fit for our data in the log$L_{2400}$--log$P_{\rm 408 ~MHz}$
plane yields $L_{2400} \propto P_{\rm 408 ~MHz}^{0.52\pm0.10}$, with a
dispersion in blue/UV magnitudes of $\sim$ 0.9, for quasars in the
redshift range $0.6-1.9$. The best-fitting slope is very similar to
that obtained by Serjeant et al. (1998), with $L_{\rm B} \propto
P_{\rm 408 ~MHz}^{0.6\pm0.1}$, although these authors give a larger
dispersion, $\sim$1.6 mag.

In the light of an intrinsic correlation between radio power and
blue/UV luminosity we can investigate in more detail the relations
between $\alpha_{ \rm UV}$, $z$, $P_{408}$ and $L_{2400}$ reported in
sections 6.4 and 6.5. The lack of an $\alpha_{\rm UV}-P_{408}$ trend
is still acceptable, in spite of an $\alpha_{\rm UV}-L_{2400}$
correlation, since the latter is weak. The trends $P_{408}-z$ and
$L_{2400}-z$ are likely related through the radio-optical correlation.  

If the correlation $L_{2400}-z$ was predominantly intrinsic, rather
than due to selection effects, the three correlations $L_{2400}-z$,
$L_{2400}-\alpha_{\rm UV}$ and $\alpha_{\rm UV}-z$ could be explained
with AD models with constant black hole mass and $\dot{m}$ increasing
with redshift (see Section 6.5). The correlation $L_{2400}-P_{408}$
would then imply that the $P_{408}-z$ trend is, at least in part,
intrinsic. If the trend $P_{408}-z$ was due only to a selection
effect, preventing the detection of low power sources at high $z$, 
the trend $L_{2400}-z$ could be also the result of this
effect, rather than cosmic evolution.  In this case one of the
correlations $\alpha_{\rm UV}-z$ or $\alpha_{\rm UV}-L_{2400}$ could
be induced by the other, in combination with the $L_{2400}-z$ trend.
According to the models by Wandel \& Petrosian (1988), a correlation
$L_{2400}-\alpha_{\rm UV}$ in the observed sense would arise naturally
if, keeping a constant black hole mass, the parameter $\dot{m}$
varies. 
We cannot exclude however the reverse interpretation, i.e. that the
intrinsic trend is for $\alpha_{\rm UV}-z$ and $\alpha_{\rm UV}$ is 
not directly dependent on blue/UV luminosity or radio power.
O'Brien et al. (1988) found from a joint regression analysis for the
three parameters that the dominant correlation for their data was
between $\alpha_{\rm UV}$ and $z$.

\section {Conclusions}
 
In this work we present optical photometry of a sample of radio
quasars in the redshift range $z=0.4-2.8$, around 80 per cent 
complete, aimed at studying their spectral energy distribution 
in the blue/UV range. 
The $U-R$, $B-R$ and $U-V$ colours do not vary substantially
with redshift, and the average values are $\langle
U-R\rangle=-0.08$ with rms 0.45, $\langle B-R\rangle=0.64$ with
rms 0.27 and $\langle U-V\rangle=-0.38$ with rms 0.42. Two quasars 
at $z$=0.50 and $z$=1.12 stand out for being particularly red, with $U-R >1$.

Power-law fits to the SEDs of the quasars with available photometry in
the four bands yield spectral indices ranging from $\sim 0.4$ to $\sim
-1.7$. The distribution of slopes is asymmetric, with a tail to steep
spectral indices. Excluding the sources in the tail, the distribution
is well modelled as a gaussian with mean and dispersion of --0.21 and
0.34 respectively.

Composite SEDs nomalized at various wavelengths were constructed from
the sample (exluding the two red quasars), and the overall shape of
the composites was found to be very similar for the different
normalizations. The only emission feature revealed from the composites
was the C{\sc iv}$\lambda$1549 line. This result is in agreement with
the expectations from the EW measurements of broad emission lines and
Fe{\sc ii} bumps of steep-spectrum radio quasars, which predict, for
the bandwidths and redshifts in our work, the largest contribution for
the C{\sc iv} line. For other emission features, like Mg{\sc ii},
C{\sc iii}], and the Fe{\sc ii} bumps at 2250--2650 and 3100-3800 \AA,
the expected average contributions to the broad band fluxes are below
ten per cent. 
The composite SEDs show a clear steepening towards high
frequencies at around 3000 \AA\ which cannot be explained by line
contamination and most likely reflect a 
trend of the continuum. Parameterizing the SEDs as power laws 
we obtained an average $\langle \alpha_{\rm
blue}\rangle= 0.11 \pm 0.16$ for the range 4500--3000 \AA\ and
$\langle \alpha_{\rm UV} \rangle = -0.66 \pm 0.15$ for 2600--1700 \AA.

Separating the quasar sample in two redshift bins, with the cut at
$z=1.2$, a better agreement was found between the values of
$\alpha_{\rm blue}$ and $\alpha_{\rm UV}$ obtained for the different
normalizations, and a hardening of $\alpha_{\rm UV}$ with redshift
emerged, with $\langle \alpha_{\rm UV}
\rangle =-0.87 \pm 0.20$ for $z<1.2$ and  $\langle \alpha_{\rm UV}
\rangle =-0.48 \pm 0.12$ for $z>1.2$. 
The average spectral index $\alpha_{\rm blue}$ for the low redshift
quasars is $\langle \alpha_{\rm blue}\rangle =0.21 \pm 0.16$,
therefore these quasars show a steepening from the blue to the UV
range $\langle\alpha_{\rm blue}\rangle- \langle\alpha_{\rm
UV}\rangle=1.08\pm0.26$. The composite SEDs for the high redshift
quasars do not cover the region above 3000 \AA, therefore the presence
of a similar break for these quasars could not be analysed from the
present data.

Correlations were also found between luminosity at 2400 \AA\ and
redshift and between radio power and redshift. Separating the quasar
sample in two bins of low and high luminosity at 2400 \AA\, a trend
was found between $\alpha_{\rm UV}$ and $L_{2400}$ (89 per cent
significant).  The quasar sample shows also an intrinsic correlation
between $L_{2400}$ and $P_{408}$ ($r_s=0.62$ and $P>99.9999$), similar
to that recently reported by Serjeant et al. (1998) for a sample of
steep-spectrum quasars.

The observed trends $L_{2400}-\alpha_{\rm UV}$, $L_{2400}-z$ and
$\alpha_{\rm UV}-z$ appear to be consistent with the predictions from
AD models for the case of constant black hole mass and accretion rate
increasing with $z$ (Wandel \& Petrosian 1988), with $M_{\rm BH}
\simeq 10^{8.2-8.5} M_\odot$, and accretion rates (in units of the Eddington value) ranging from $\dot{m} \simeq$ 1 for $z=0.6-1.2$ to $\dot{m} \simeq$ 10 
for $z=1.2-1.9$.  An alternative interpretation is that the $P_{408}-z$ 
and $L_{2400}-z$ trends arise predominantly from a selection effect, due to 
the radio flux limits of the sample. In this case one of the correlations, 
$\alpha_{\rm UV}-z$ or $\alpha_{\rm UV}-L_{2400}$, could be induced by the 
other, in combination with the $L_{2400}-z$ trend. The observed 
$\alpha_{\rm UV}-L_{2400}$ correlation is consistent with the predictions 
from the AD models by Wandel \& Petrosian assuming
a constant black hole mass and a range of accretion rates, although in this case the accretion rate does not need to be
correlated with $z$.  We cannot exclude however that the intrinsic
correlation is $\alpha_{\rm UV}$--$z$, with
$\alpha_{\rm UV}$ not being physically related to the blue/UV
luminosity.

\section {Acknowledgements}

The 1.0-m JKT telescope is operated on the island of La Palma by the Isaac Newton Group in the
spanish Observatorio del Roque de los Muchachos of the Instituto de
Astrof\'\i sica de Canarias. The 2.2-m telescope, at the
Centro-Astron\'omico Hispano-Alem\'an, Calar Alto, is operated by the
Max-Planck-Institute for Astronomy, Heidelberg, jointly with the
spanish Comisi\'on Nacional de Astronom\'\i a. RC, JIGS and SFS
acknowledge financial support by the DGES under project PB95-0122 and
by the Comisi\'on Mixta Caja Cantabria-Universidad de Cantabria.
SFS wants to thank the FPU/FPI program of the Spanish MEC for a fellowship.


\begin{thebibliography}{}

\bibitem[Baker \& Hunstead \ 1995]{ba95} Baker J.C. and Hunstead R.W.,
1995, ApJ , 452, L95. Erratum: ApJ, 468, L131 

\bibitem[Barvainis 1993]{bar93} Barvainis R., 1993, ApJ, 412, 513




\bibitem[Boroson \& Green \ 1992]{bg92} Boronson T.A., Green R.F., 
1992, ApJS 80, 109

\bibitem[Browne \& Wright \ 1985]{bw85} Browne I.W.A., Wright A.E., 
1985, MNRAS, 213, 97

\bibitem[Burnstein \& Heiles \ 1982]{burn82} Burnstein D., Heiles C.,
1982, AJ, 87, 1165

\bibitem[Carballo et al. \ 1998]{car98} Carballo R., S\'anchez S.F.,
Gonz\'alez-Serrano J.I., Vigotti M., Benn C., 1998, AJ, 115, 1234

\bibitem[Czerny \& Elvis \ 1987]{ce87} Czerny B., Elvis M., 
1987, ApJ, 321, 305

\bibitem[Cristiani \& Vio \ 1990]{Cris} Cristiani S., Vio R., 1990, A\&A, 
227, 385

\bibitem[Cristiani et al 1995]{Cris95} Cristiani S., et al. 1995, A\&AS,
112, 347

\bibitem[Elvis et al. 1994]{elv94} Elvis M., et al. 1994, ApJSS, 95, 1 

\bibitem[Ficarra \ 1985]{fi95} Ficarra A., Grueff G., Tomassetti G., 1985, 
A\&AS 59, 255

\bibitem[Francis et al \ 1991]{fra91} Francis P.J., Hewett P.C., Foltz
C.B., Chaffee F.H., Weymann R.J., Morris S.L., 1991, ApJ, 373, 465

\bibitem[Green \ 1998]{g98} Green P.J., 1988, 
ApJ, 498, 170

\bibitem[Hewett et al. 1991]{hew91} Hewett P.C., Foltz C.B., Chaffee
F.H., Francis P.J., Weyman R.J., Morris S.L., Anderson S.F., MacAlpine
G.M., 1991, AJ, 101, 1121

\bibitem[Johnson \ 1966 ]{j66} Johnson H.L., 1966, ARAA 4, 193

\bibitem[Landolt \ 1992 ]{land92} Landolt A.U., 1992, AJ, 104, 340

%\bibitem[Large et al. \ 1991 ]{lar91} Large M.I., Mills B.Y., Little A.G., 
%Crawford D.F., Sutton J.M., 1981, MNRAS, 194, 693

%\bibitem[Laing et al. \ 1983 ]{lai83} Laing R.A., Riley J.M., Longair M.S., 
%1983, MNRAS, 204, 151

\bibitem[La Franca et al 1992]{franc92} La Franca F., Cristiani S.,
Barbieri C., 1992, AJ, 103, 1062

\bibitem[Malkan \ 1983]{m83} Malkan M.A., 1983, ApJ, 268, 582

\bibitem[Natali et al. \ 1998]{Nat98} Natali F., Giallongo E.,
Cristiani S., La Franca F., 1998, AJ, 115, 397

\bibitem[O'Brien \ 1988]{o88} O'Brien P.T., Gondhalekar P.M., Wilson R., 
 1988, MNRAS, 233, 801

\bibitem[Rieke \& Lebofsky \ 1985]{rl85} Rieke G.H., Lebofsky M.J., 
1985, ApJ, 288, 618 

\bibitem[Sanders et al. 1989]{s89} Sanders D.B., Phinney E.S., Neugebauer G., 
Soifer B.T., Mathews K., 1989, ApJ, 347, 29

%\bibitem[Schmidt \& Green 83]{sg83} Schmidt M. Green R., 1983, ApJ, 269, 352

\bibitem[Serjeant et al. 1997]{sj97} Serjeant S., Rawlings S., Maddox
S.J., Baker J.C., Clements D., Lacy M., Lilje P.B., 1998, MNRAS,
294, 494

\bibitem[Vigotti et al. \ 1989]{vig89} Vigotti M., Grueff G., Perley
R., Clark B.G., Bridle A.H., 1989, AJ, 98, 419

\bibitem[Vigotti et al. \ 1997]{vig97} Vigotti M., Vettolani G.V.,
Merighi R., Lahulla J.F., Pedani M., 1997, A\&AS, 123, 1 

\bibitem[Wandel Petrosian  \ 1988]{wp88} Wandel A., Petrosian V.,
1988, ApJ, 329, L11
 
\bibitem[Wamsteker \ 1981]{w81} Wamsteker, W., 1981, A\&A, 97, 329

\bibitem[Willot et al  \ 1998]{wp98} Willott C.J., Rawlings S., Blundell K.M., 
Lacy M., 1998, MNRAS, 300, 625

\bibitem[Wills et al. \ 1985]{w85} Wills B.J., Netzer H., Wills D., 
1985, ApJ, 288, 94

\bibitem[Zheng et al. \ 1997]{z97} Zheng W., Kriss G.A., Telfer R.C., 
Grimes J.P., Davidsen A.F., 1997, 
ApJ, 475, 469

%\bibitem[Zombeck \ 1990]{zom90} Zombeck M.V., 1990, In {\it 
%Handbock of Space Astronomy \& Astrophysics}, Cambridge University Press.

\end{thebibliography}
\end{document}